\documentclass{aa}
\usepackage{graphics}
\usepackage{graphicx}
\usepackage{amssymb}
\begin{document}

\def\srm{$\sigma_{\rm RM}$}
\def\rmm{$\langle{\rm RM}\rangle$}
\def\absrmm{$\arrowvert \langle {\rm RM} \rangle  \arrowvert$}
\def\bm{$\langle\mathbf B\rangle$}

\title{Magnetic Fields and Faraday Rotation in Clusters of Galaxies}

\author{M. Murgia \inst{1,2} \and F. Govoni \inst{1,3}  \and
 L. Feretti \inst{1} \and G. Giovannini \inst{1,3} \and D. Dallacasa
 \inst{1,3} \and R. Fanti \inst{1,4} \and G. B. Taylor  \inst{5} \and K. Dolag \inst{6}}

\institute{
Istituto di Radioastronomia del CNR, Via Gobetti 101, I-40129 Bologna, Italy
\and
INAF$-$Osservatorio Astronomico di Cagliari, Loc. Poggio dei Pini, Strada 54,
I-09012 Capoterra (CA), Italy
\and
Dipartimento di Astronomia, Universit{\`a} di Bologna, Via Ranzani 1,
I-40127 Bologna, Italy
\and
Dipartimento di Fisica, Universit{\`a} di Bologna, Via Irnerio 46,
 I-40126 Bologna, Italy
\and
National Radio Astronomy Observatory, Socorro, NM 87801, USA
\and
Dipartimento di Astronomia, Universit{\`a} di Padova, vicolo dell'Osservatorio 5, 
35122 Padova, Italy
}

\date{Received; Accepted}

\abstract{

We present a numerical approach to investigate the relationship
between magnetic fields and Faraday rotation effects in clusters of
galaxies.  We can infer the structure and strength of
intra-cluster magnetic fields by comparing our simulations with the
observed polarization properties of extended cluster radio sources
such as radio galaxies and halos. We find the observations
require a magnetic field which fluctuates over a wide range of spatial
scales (at least one order of magnitude). If several polarized
radio sources are located at different projected positions in a galaxy
cluster, as is the case for A119, detailed Faraday rotation
images allow us to constrain both the magnetic field strength
and the slope of the power spectrum. 
Our results show that the standard analytic expressions applied
in the literature overestimate the cluster magnetic field
strengths by a factor of $\sim$2. We investigate the
possible effects of our models on beam depolarization of radio sources
whose radiation traverses the magnetized intracluster
medium. Finally, we point out that radio halos may provide
important information about the spatial power spectrum of the magnetic
field fluctuations on large scales. In particular, different values of
the index of the power spectrum produce very different
total intensity and polarization brightness distributions.

}

\offprints{M. Murgia, murgia@ira.cnr.it}

\titlerunning{Magnetic Fields and Faraday Rotation in Clusters of Galaxies.}
\authorrunning{M. Murgia et al.}
\maketitle

\section{Introduction}

The existence of cluster scale magnetic fields has been demonstrated 
 in different ways. Some clusters of galaxies exhibit
non-thermal synchrotron radio halos which extend up to Megaparsec
scales. These diffuse radio sources constitute direct evidence
for large scale magnetic fields associated with the intracluster
medium. Indirect evidence for cluster scale magnetic fields are
provided by the observation of Faraday rotation in radio sources
located inside or behind galaxy clusters. For a review of these
and other techniques see Carilli \& Taylor (2002).

A crucial issue concerns the effective strength and structure of 
 cluster magnetic fields since different methods 
of analysis give somewhat discrepant estimates for the field strengths.
 Due to their low surface brightness, radio halos
 have been studied so far with low spatial resolution. This prevents a
 detailed observation of the small scale magnetic field 
 geometry and intensity. 
However, using minimum energy assumptions, it is possible to estimate 
an equipartition magnetic field strength averaged over the 
entire halo volume. 
These estimates give equipartition magnetic field strengths of
$\simeq$0.1 to 1 $\mu$G (e.g. Feretti \& Giovannini 1996, Govoni et al. 2001a,
Bacchi et al. 2003).
In a few cases, clusters containing a radio halo 
show a hard X-ray excess emission. This emission could be interpreted in terms of 
Inverse Compton scattering of the cosmic microwave background photons with the 
relativistic electrons responsible for the radio halo emission. In this case,
the measures of the magnetic field strength (e.g. Fusco-Femiano et al. 1999,
Rephaeli et al. 1999) 
inferred from the
ratio of the radio to X-ray luminosities are consistent with the 
equipartition estimates. Because of the lack of spatial information 
 of these X-ray observations also in this case we obtain a field
 strength averaged over a large region.
Indirect measurements of the magnetic field
strength can also be determined in conjunction with X-ray observations 
of the hot gas, through the study of the Faraday Rotation Measure (RM) of 
radio sources located inside or behind clusters. 
By using simple analytical approaches, relatively strong magnetic fields,
from $\sim$5 $\mu$G up to the values of 30 $\mu$G 
have been found in cooling flow  
clusters (e.g., 3C~295, Perley \& Taylor 1991 and 
Allen et al. 2001; Hydra A, Taylor \& Perley 1993) 
where extremely high RMs have been revealed.
On the other hand, significant magnetic fields have also been 
detected in clusters without cooling flows: the RM measurements
of polarized radio sources through the hot intracluster medium
leads to a magnetic field of $2-8$ $\mu$G with
 patchy structures on small scales in the range $2-15$ kpc 
(e.g. Feretti et al. 1995, Feretti et al. 1999a, Govoni et al. 2001b, 
Taylor et al. 2001, Eilek \& Owen 2002). 
From a statistical study of several clusters, 
Clarke et al. (2001) obtained magnetic fields
of $4-8$ $\mu$G, in agreement with the above findings.

The magnetic field strength obtained by RM studies is therefore higher
than the value derived from the radio halo data and from
Inverse-Compton X-ray studies. However, as pointed out by
Carilli \& Taylor (2002 and references therein), all the aforementioned 
techniques are based on several assumptions. For
example, the observed RMs have been interpreted until now in terms
of simple analytical models which consider single-scale magnetic
fields, while equipartition calculations in radio halos assume
spatially uniform magnetic fields.

Newman et al. (2002) demonstrated that the assumption of 
a single-scale magnetic field 
leads to an overestimation of the magnetic field strength calculated through
RM studies. Moreover, detailed magneto-hydrodynamic cosmological simulations
(Dolag et al. 2002) suggested that the cluster magnetic fields may 
span a wide range of spatial scales with a strength that decreases with 
distance from the cluster center.

 Finally, It has been pointed out that in some cases a radio source 
could compress the gas and fields in the ICM to produce local RM enhancements, 
thus leading to overestimates of the derived ICM magnetic field strength (e.g. 
Feretti el al. 1995, Rudnick \& Blundell 2003). No investigation in this respect has been attempted 
in this work.

So far, very little attention has been given in the literature to the determination of the power
 spectrum of the intra-cluster magnetic field fluctuations. Only recently  En{\ss}lin \& Vogt (2003) and Vogt \& En{\ss}lin
 (2003) by using a new semi-analytic technique showed that, for those cluster sources for which a 
very detailed RM image is available, the magnetic field power spectrum can be estimated.

The aim of our work is to provide an alternative numerical approach for investigating the strength and
structure of cluster magnetic fields through Monte Carlo simulations. The simulations results are compared
 to the data through an interactive technique. The algorithms are assembled into a software package named FARADAY. 

In particular, we will use FARADAY to study the power spectrum of the 
intra-cluster magnetic field from two different but complementary points of view.
First, we show that both the mean and the dispersion of the RM can be used (and therefore should be used) 
 to constrain not only the strength but also the power spectrum slope of the magnetic field fluctuations. 
In a consistent way, our simulations can be also employed to derive the expected de-polarization of   
 cluster sources, such as radio galaxies and large scale halos, thus improving the estimates based on the RM analysis.

Throughout this paper we will assume a simple power law form for the
magnetic field power spectrum and compare the observed cluster
radio source polarization and RM images with the simulated values
over regions of comparable size. The reader should be aware that many
results presented in this work are influenced by the
form assumed for the power spectrum and by the limitations of the
numerical simulations. However, by considering the magnetic field strength as radial
dependent and by modeling the magnetic field fluctuations within a
multi-scale length approach we made a new step towards more realistic cluster magnetic fields models.
As we will show, modeling of magnetic fields as those we propose are necessary to 
interpret Faraday rotation effects in galaxy clusters and should be seen as a starting point for a new line of investigations.

This paper is organized as follows:
in Sect.~2 we give a brief description of the FARADAY tool.
In Sect.~3 we discuss the three-dimensional multi-scale cluster magnetic fields 
models as implemented in the program and used in our simulations.
In Sect.~4 we simulate the Rotation Measure.
In Sect.~5 we simulate the beam depolarization of cluster radio galaxies and the morphology and depolarization 
of the radio halo emission.
In Sect.~6 we present some comparisons of simulations with real data.
Some discussion and conclusions are presented in Sect.~7.

In this work we adopt a Hubble constant H$_{\rm 0}$=50~km~s$^{-1}$Mpc$^{-1}$
and a deceleration parameter $q_{\rm 0}=0.5$. 

\section{Faraday rotation in clusters of galaxies and the FARADAY tool}

The polarized synchrotron radiation incoming from radio sources
located inside or behind a galaxy cluster, experiences a rotation of the
plane of polarization as it passes through the magnetized and ionized 
intra-cluster medium:

\begin{equation}
\Psi _{\rm Obs}(\nu)=\Psi _{\rm Int}+(c/\nu) ^2 \times {\rm RM}
\label{psi}
\end{equation}

\noindent
where $\Psi _{\rm Obs}(\nu)$ is the observed polarization angle
at a frequency $\nu$ and $\Psi _{\rm Int}$ is the 
intrinsic polarization angle. 

The RM is related to the thermal electron density, $n_{\rm e}$, 
and magnetic field along the line-of-sight, $B_{\|}$, 
 through the cluster by the equation:
\begin{equation}
{\rm RM} = 812\int\limits_0^L n_{\rm e} B_{\|} {\rm d}l ~~~{\rm rad~m}^{-2}
\label{rm}
\end{equation}
where $B_{\|}$ is measured in $\mu$G, $n_{\rm e}$
in cm$^{-3}$ and $L$ is the depth of the screen in kpc.   

The position angle of the polarization plane is an observable
quantity, therefore, RM images of radio sources can be constructed, by
a linear fit to Eq. \ref{psi}.  The observed degree of
polarization intensity, $P_{\rm Obs}(\nu)$, can be significantly lower
with respect to the intrinsic value, $P_{\rm Int}$, if differential
Faraday rotation occurs within the observing beam. The observable data, RM
 and (de)polarization, together
 with  a model for the density distribution of the 
 X-ray emitting hot gas can provide, in principle, 
important information on the cluster magnetic field strength and structure.
 However, there are  a number of practical difficulties that
 complicate the applicability of this method. In particular, due to
 the random character of the cluster magnetic fields,   
Eq.~\ref{rm} is not analytically solvable even for very simple
 distributions of the thermal electrons. Moreover, we will always
 observe a projection of the random field onto the two-dimensional
 sky, so we would need to de-project the power spectrum and
 the auto-correlation function radio source RM images in order
 to reconstruct the magnetic field configuration. This may
 be a non-trivial process (En{\ss}lin \& Vogt 2003).

For this purpose we developed the simulation tool FARADAY which allows 
an interactive, numerical approach for investigating cluster 
magnetic fields. 

 The thermal electron gas density of the screen is generally not known, 
and we then assume a standard $\beta$-model distribution:

\begin{equation}
n_{\rm e}(r)=n_{\rm 0}(1+r^2/r^2_{\rm c})^{-3\beta/2}$$
\label{king}
\end{equation}

\noindent
where $r$, $n_{\rm 0}$ and $r_{\rm c}$ are the distance from the cluster center,
 the central electron density, and the cluster core radius, respectively.

By following some simple steps it is possible 
to compute the effects of a cluster magnetic field model 
on the  polarization properties 
of sources located inside or behind a cluster.
Given a three-dimensional magnetic field model and  
 the density distribution of the intra-cluster gas, 
 FARADAY calculates the RM image by integrating
 Eq.~\ref{rm} numerically. 

The beam depolarization is expected to be particularly strong
 if the radio source is observed  with a beam larger than the minimum magnetic
field scale length. This effect can be studied with FARADAY as well.
Starting from the intrinsic polarization images of $P_{\rm Int}$ and
 $\Psi_{\rm Int}$, FARADAY can simulate the expected beam depolarization
 as a function of wavelength.
 One of the key feature of FARADAY is that it allows us to study the
effect of magnetic field models that cannot be investigated analytically.
In this work we use FARADAY to investigate the
effects of a turbulent and isotropic magnetic fields which
fluctuate over a range of spatial scales.

\section{A multi-scale magnetic field model}

Detailed images of the polarized emission  are now available for a significant
number of radio sources seen in galaxy clusters and indicate that the
 RM fluctuates down to linear scales as low as 10 kpc or less (e.g. Carilli \& Taylor 2002 and references therein).
If interpreted in terms of an external Faraday screen, the dispersion, $\sigma_{\rm RM}$, and mean, $\langle {\rm RM}
\rangle$, of the RM fluctuations can be used to constrain the cluster magnetic field strength.

The effect of Faraday rotation from a tangled magnetic field has been
analyzed by several authors (e.g. Lawler \& Dennison 1982, Tribble
1991, Feretti et al. 1995 and Felten 1996).  In the simplest ideal
case, the screen is made of cells of uniform size, electron density
and magnetic field strength, but with a field orientation at random
angles in each cell. The observed RM along any given line of sight is
then generated by a random walk process involving a large
number of cells of size $\Lambda_{\rm c}$.  The distribution of the RM
is Gaussian with zero mean, and a variance given by:
\begin{equation}
 \sigma_{\rm RM}^{2}= \langle {\rm RM^{2}} \rangle = 812^{2} \Lambda_{\rm c} \int ( n_{\rm e} B_{\|})^{2}
{\rm d}l~.
\label{sigmarndwalk}
\end{equation}

As we will show in Sect.\ref{monoscale}, since the density profile of the ionized gas can be obtained by X-ray
observations, the intergalactic magnetic field strength can be estimated by
measuring $\sigma_{\rm RM}$ from spatially resolved RM images of
 radio sources in the cluster if $\Lambda_{\rm c}$ is known.

In many cases the observed RM distributions are nearly Gaussians,
suggesting an isotropic distribution of the field component along the
line-of-sight. However, many RM distributions show clear evidence for
a non-zero \rmm~if averaged over areas comparable with the radio
source size  which is typically  50 $\times$ 50 kpc$^2$.
This same property is observed for clusters located at
high galactic latitudes and in general cannot be entirely attributed
to the contribution from our own Milky Way Galaxy.  These \rmm~offsets
are more likely due to fluctuations of the cluster magnetic fields on
scales greater than 50$-$100 kpc, i.e. considerably larger of those
responsible for the RM dispersion.
The random magnetic field must therefore be both tangled on sufficiently
small scales in order to produce the smallest structures observed in
the RM images and also fluctuate on scales one, or even two,
orders of magnitude larger.  For this reason, it is necessary to
consider cluster magnetic field models where both small and large
scales structures coexist.  To accomplish this we simulated in a
cubical box a three-dimensional multi-scale magnetic field.

Following the approach proposed by Tribble (1991), the field is constructed by 
selecting a power spectrum for the  vector potential $\mathbf A$ and choosing
 the Fourier components $\mathbf {\tilde A}(\mathbf k)$ accordingly.

For all grid points in the wave-numbers space the amplitude and phase, respectively $A$ and $\phi$, 
 of each component of $\mathbf {\tilde A}(\mathbf k)$ are randomly drawn from the distribution:
\begin{equation}
P(A,\phi)dAd\phi=\frac{A}{2\pi~\arrowvert A_{k}\arrowvert ^2}\exp\left(-\frac{A^2}{2~\arrowvert A_{k}\arrowvert ^2}\right)dAd\phi
\label{pafi}
\end{equation}
i.e.  $A$ is extracted from a Rayleigh distribution and $\phi$ is uniformly 
distributed between 0 and $2\pi$. 
 
We adopted a power-law power spectrum\footnote{Note that throughout this 
paper the power spectra are expressed as  
 vectorial forms in $k$-space. The one-dimensional forms can be obtained by multiplying by $4\pi k^{2}$ and $2\pi k$
 respectively the three and two-dimensional power spectra.

}, $\arrowvert A_{k}\arrowvert ^2$, of the  form:

\begin{equation}
\arrowvert A_{k}\arrowvert ^2 \propto k^{-\zeta}.
\label{powera}
\end{equation}

The magnetic field:
\begin{equation}
 {\mathbf {\tilde B}}(\mathbf k)=i{\mathbf k} \times \mathbf {\tilde A}(\mathbf k)
\label{powerb}
\end{equation}
is then transformed back into
 the real space using a three-dimensional Fast Fourier
Transform (Press et al. 1986). This automatically ensures a tangled, divergence free,
multi-scale magnetic field model whose components have a power spectrum which follows:

\begin{equation}
|B_k|^2= C_{n}^{2} k^{-n}
\label{bpower}
\end{equation}
where $n=\zeta-2$ while $C_{n}^{2}$ is the power spectrum normalization.

The power spectrum (\ref{bpower}) represents the magnetic field energy 
(erg cm$^3$)
associated with each wave-number. 

The assumed form 
for the  vector potential results in
 a magnetic field whose components, $B_{i}$, have a Gaussian distribution with
 $\langle B_{i}\rangle = 0$ and $\sigma_{B_{i}}^{2}=\langle B_{i}^2\rangle$. The distribution of $B$ itself is therefore a Maxwellian
 with $\langle B\rangle = 2\sqrt{\frac{2}{\pi}}~\sigma_{B_{i}}$ and $\sigma_{B}=\sqrt{\frac{3\pi-8}{\pi}}~\sigma_{B_{i}}$.

From the Parseval's Theorem, $\int \arrowvert B_{i} \arrowvert ^{2} dV= \int|B_k|^2 d^{3}k$, it follows that
 the magnetic field energy density averaged over the volume, $V$, is proportional to

\begin{equation}
\langle B\rangle^{2}\propto \frac{C_{n}^{2}}{V}\cdot\ln(k_{\rm max}/k_{\rm min})
\label{normbmedio}
\end{equation}

for $n=3$, and 

\begin{equation}
\langle B\rangle^{2}\propto \frac{C_{n}^{2}}{V}\cdot \frac{k_{\rm max}^{3-n}-k_{\rm min}^{3-n}}{3-n}
\label{normbmedio}
\end{equation}

for $n\neq3$. Where the wave-numbers $k_{\rm min}$ and $k_{\rm max}$  
correspond, respectively, to the minimum and maximum spatial scale 
of the magnetic field.  

\subsection{A magnetic field strength with a radial decrease}
 
Important clues about the radial gradient of the magnetic field
strength in clusters of galaxies are provided by observations of
clusters hosting a radio halo and by magneto-hydrodynamic cosmological
simulations.  The spatial correlation found in some clusters between
the X-ray and the radio halo brightness (Govoni et al. 2001c) 
suggests that thermal and
non-thermal (relativistic particles and magnetic fields) energy
densities could have the same radial scaling.
In the  radio halo of the Coma cluster (Giovannini et al. 1993) 
a steepening of the synchrotron spectrum is observed from the cluster center
outward.  Such a spectral behavior is expected in some halo formation
models which consider a radial decrease of the cluster magnetic
field strength (e.g. Brunetti et al. 2002). Moreover, detailed
magneto-hydrodynamic cosmological simulations (Dolag et al. 2002) lead
to magnetic fields whose strength decreases  with the distance 
from the cluster center.

Therefore in this work we considered a magnetic field whose strength decreases
 from the cluster center according to:
\begin{equation}
\langle\mathbf B\rangle(r)=\langle\mathbf B\rangle_{\rm 0}\cdot(1+r^{2}/r_{\rm c}^{2})^{-{\bf \frac{3}{2}}\mu}
\label{br}
\end{equation}
where $\langle\mathbf B\rangle_{\rm 0}$ is the mean magnetic field at the cluster center. 
Given the electron density profile in Eq.~\ref{king}, we have $\langle\mathbf B\rangle \propto n_{\rm e}^{\mu/\beta}$. Therefore 
the case $\mu=\beta/2$ corresponds to a magnetic
 field whose energy density decreases from the cluster center as the gas energy density
 while $\mu=2\beta/3$ corresponds to a magnetic field frozen in the matter.

By combining RM and X-ray data it is possible to obtain an estimate of the 
index $\mu$ (Dolag et al. 2001). 
The X-ray surface brightness, $S_{\rm x}$, in a case of the $\beta$-model (Eq.~\ref{king}), is proportional to 
\begin{equation}
S_{\rm x} \propto (1+r_{\perp}^{2}/r_{\rm c}^{2})^{-3\beta+{\bf \frac{1}{2}}}
\end{equation}
while, by substituting Eq.~\ref{br} in Eq.~\ref{sigmarndwalk}, the RM is proportional to
\begin{equation}
\langle\rm RM^2\rangle^{1/2} \propto (1+r_{\perp}^{2}/r_{\rm c}^{2})^{\bf {-\frac{3}{2}(\beta+\mu)+\frac{1}{4}}}~.
\end{equation}
where $r_{\perp}$ is the projected distance from the cluster center. 
By comparing the two line-of-sight integrals, one finds that the index $\mu$ is related to the 
slope, $\alpha$, of the correlation $\langle\rm RM^2\rangle^{1/2}\propto S_{\rm x}^{\alpha}$ and to the density index, $\beta$, 
through
\begin{equation}
\mu=(2\alpha-1)\cdot(\beta-1/6)~.
\label{alpha}
\end{equation}
Thus, for a constant magnetic field ($\mu=0$)  
the slope of the RM$-S_{\rm x}$ correlation
 should be $\alpha=0.5$ while a steeper slope would imply $\mu > 0$.  

Formally, a decreasing magnetic field strength, as in Eq.~\ref{br},
should be achieved by convolving the  vector potential components with
the shaping profile in the Fourier space.  A more practical, although
less rigorous, way to obtain this, consists in shaping the magnetic
field to the desired profile directly in real space. We checked
that in our cases the two approaches give minimal differences and
for simplicity we adopted the latter.  It should be noted that the magnetic
field power spectrum in both cases is given by the convolution of the
power-law in Eq.~\ref{bpower} with the Fourier transform of the
shaping function. However, since the core radius of the shaping
function is a considerable fraction of the simulated volume the
convolution does not significantly alter the form of the power-law in
Eq.~\ref{bpower}.

\section{Simulated Rotation Measures}
\label{rmsim}

Using the cluster magnetic field model described in Sect.~3, we
studied the behavior of the RM for different values of the spectral
index of the power spectrum in the range from $n=2$ to $n=4$.  The
magnetic energy density scales as $d u_{\rm B}/dk\propto k^2|B_k|^2$,
thus $n=3$ implies that the field is scale-invariant, i.e. the energy
density per logarithmic wavenumber interval is constant.  It follows
that for $n<3$ the magnetic energy density is larger on the smaller
scales while for $n>3$ most of the magnetic energy density is on the
larger scales.  We now explore all these possibilities.

\subsection{Simulated RM images}
\label{rmimages}

\begin{figure*}
\begin{center}
\includegraphics[height=20cm]{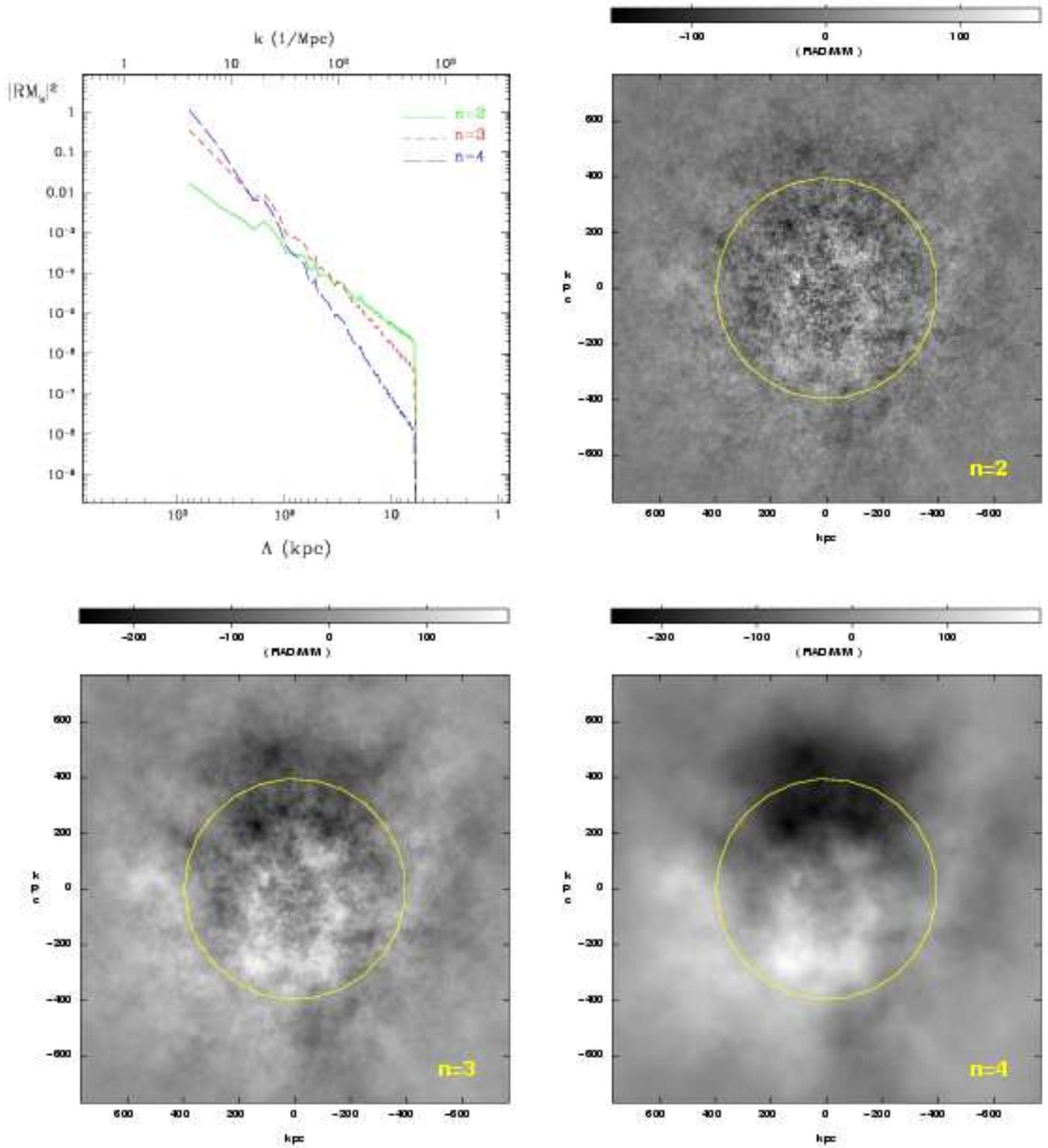}
\end{center}
\caption[]{Simulated RM images for different values of the magnetic field power 
spectrum spectral index $n$. The three power spectra are normalized to have the 
same total magnetic field energy which is distributed over the range of spatial scales
 from 6 kpc up to 770 kpc. The average field at the cluster center is \bm$_{\rm0}$ $=1 ~\mu$G and its  
energy density decreases from the cluster center according to $B^{2}\propto n_{\rm e}(r)$, where 
$n_{\rm e}(r)$ is the gas density profile.
 Each RM image shows a field of view of about $1.5\times 1.5$ Mpc while the cluster core
 radius (indicated by the circle) is 400 kpc. We performed the RM integration  
 from the cluster center up to 3.8 core radii along the line-of-sight.
The two-dimensional power spectra of the simulated RM images are shown in the top left panel.
 They have the same slope as their parent magnetic field power spectra and
 they span an equivalent range of spatial scales -- see text for details.}
\label{fig1}
\end{figure*}

\begin{figure*}[t]
\begin{center}
\includegraphics[width=18cm]{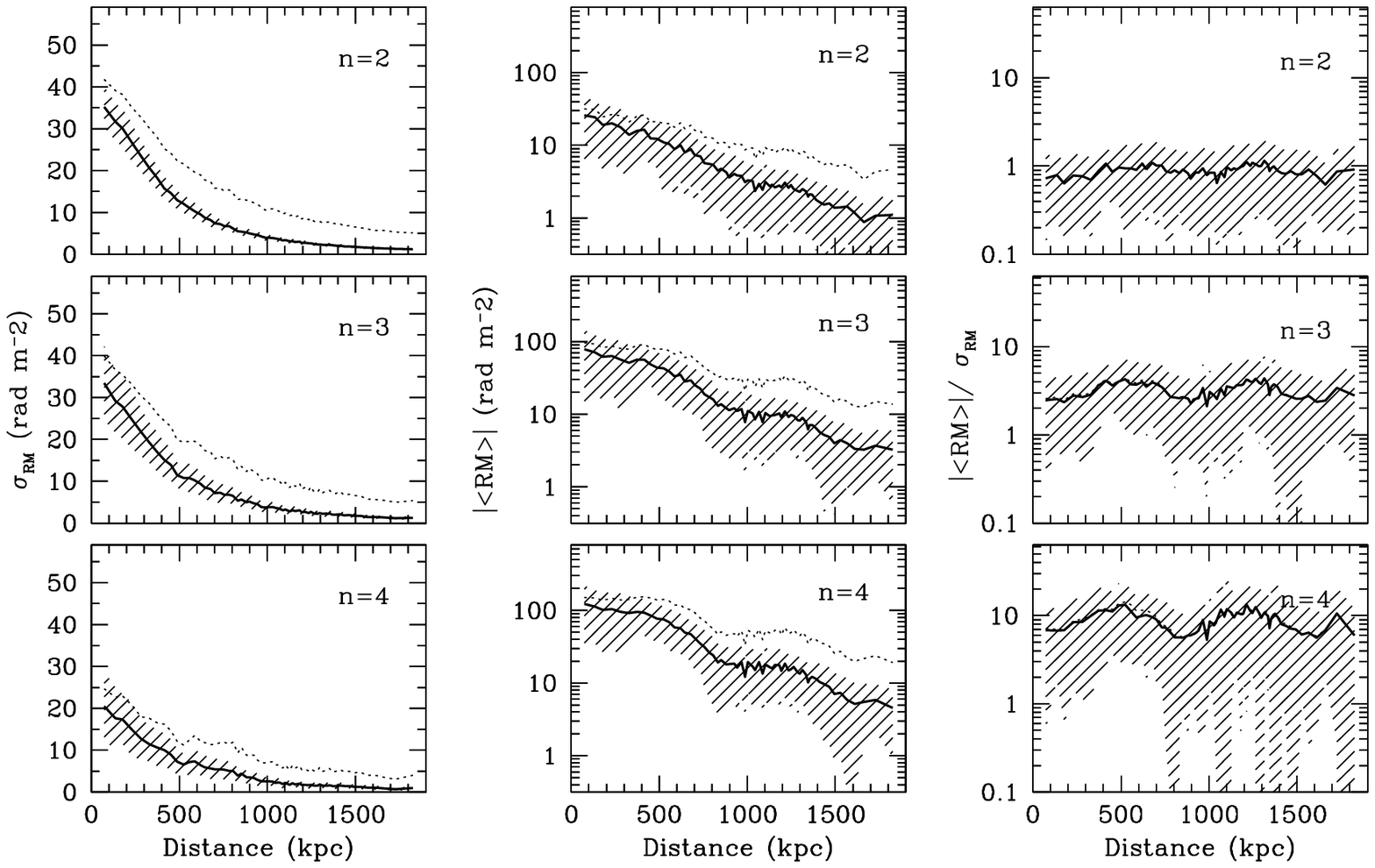}
\end{center}
\caption[]{Radial profiles obtained from the RM simulations described in {Sect.~}\ref{rmsim}. 
Left, central and right panels show the radial trends of \srm, \absrmm~and \absrmm/\srm, respectively,
 for three values of the index of the power spectrum $n$.
Solid lines correspond to an average field at the cluster center of \bm$_{\rm 0}=1 ~\mu$G whose 
energy density decreases from the cluster center according to $B^{2}\propto n_{\rm e}(r)$, where 
$n_{\rm e}(r)$ is the gas density profile. The shaded regions indicate the 1-rms statistical fluctuations due to
 the random nature of the magnetic field.
Dotted lines represent the profiles corresponding to a constant magnetic field strength of \bm$_{\rm 0}=1 ~\mu$G
 all over the cluster.
Values have been sampled in $50 \times 50$ kpc$^2$ regions.}
\label{fig2}
\end{figure*}

\begin{table}[t]
\caption{Parameters adopted in the simulations.}
\begin{center}
\begin{tabular}{ll}
\noalign{\smallskip}
\hline                  
\noalign{\smallskip}    
 Grid size              & $512^{3}$ pixels \\   
 Cellsize               & 1 pixel=3 kpc \\
\hline
 Core radius        & $r_{\rm c}$=400 kpc         \\
 Central density    & $n_{\rm 0}$=$10^{-3}$ cm$^{-3}$    \\
 Beta               & $\beta$=0.6 \\ \hline
 Central Mean magnetic field    & $\langle{\mathbf B}\rangle_{\rm 0}$=1 $\mu$G \\
 Radial profile slope                   & $\mu=0, 0.3$            \\
 Magnetic field minimum scale    & $\Lambda_{\rm min}$=6 kpc \\
 Magnetic field maximum scale    &  $\Lambda_{\rm max}$=768 kpc\\
 Power spectrum spectral index   & $n=$2, 3, 4\\
\hline
\noalign{\smallskip}
\label{simul}
\end{tabular}
\end{center}
\end{table}

\begin{figure*}[t]
\begin{center}
\includegraphics[width=18.5cm]{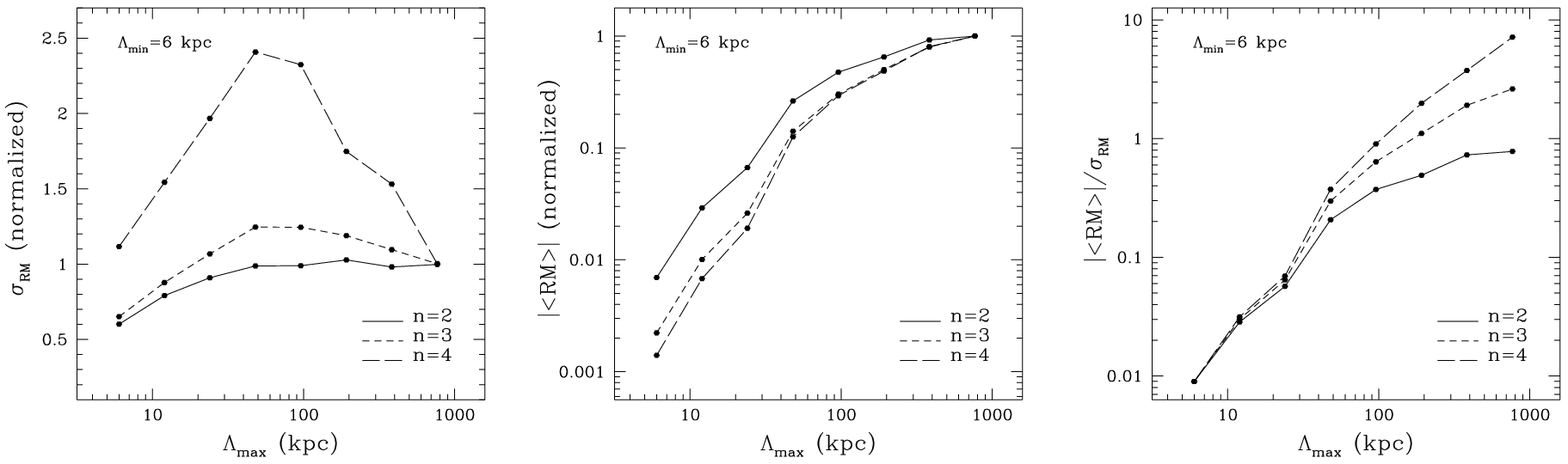}
\end{center}
\caption[] {Normalized dependence of \srm (left) and \absrmm (middle)
as a function of $\Lambda_{\rm max}$, for the three spectral indexes
$n$=2, 3 and 4. The right panel shows the ratio \absrmm/\srm~as a
function of $\Lambda_{\rm max}$. The minimum magnetic field scale has
been fixed at $\Lambda_{\rm min}=6$ kpc while \srm~and \absrmm~values
have been sampled in $50 \times 50$ kpc$^{2}$ regions.}
\label{fig3}
\end{figure*}

We simulated the expected RM images corresponding to the three different values
of the power spectrum spectral index mentioned above in the case of a cluster 
 with typical size and gas density parameters.
The computational grid, gas density and magnetic field model
parameters used for these specific simulations are listed in Tab.~\ref{simul} while the
 RM images we obtained are shown in  Fig.~1. Each RM image shows a field of view of about
$1.5\times 1.5$ Mpc while the cluster core radius (indicated by the
circles in Fig.~1) is taken  to be 400 kpc. We performed the integration of
Eq.~\ref{rm} from the cluster center up to 1.5 Mpc (i.e. 3.8 $r_{\rm
c}$) along the line-of-sight. To do this we located the cluster center in the 
 farthest side of the computational grid.
The three magnetic field power spectra are normalized to generate the same
average magnetic field energy density. The average field at the cluster center
is \bm$_{\rm 0}=1 ~\mu$G and its energy density decreases from the
cluster center as the gas energy density profile, i.e. $\mu=0.3$ (see
Eq.~\ref{br}).  This corresponds to a mean magnetic field of
\bm$\simeq 0.46~\mu$G within a spherical volume with a radius of $r=3r_{\rm c}$. 

Since the size of the computational grid is finite only a limited
range of spatial scales can be investigated.
The grid is large enough to contain a cluster of typical size while
keeping the necessary resolution to sample reasonably well the
magnetic field structure on a spatial scale\footnote{ Here we refer to 
the length $\Lambda$ as to the magnetic field reversal scale. In this way, $\Lambda$ 
corresponds to a half-wavelength, i.e. $\Lambda= 0.5\cdot(2\pi/k)$.} range from $\Lambda_{\rm
min}=6$ up to $\Lambda_{\rm max}\simeq 770$ kpc. Here we adopted the maximum range 
of spatial scales permitted by the computational grid.
Observations reveal RM fluctuations on scales as low as 10 kpc supporting this choice for $\Lambda_{\rm min}$.
The exact value of $\Lambda_{\rm max}$ is more uncertain, however the presence in some 
 cluster of galaxies of wide radio halos may indicate that at least in these cases the cluster magnetic 
field is  spread over Mpc distances. Simulations with a different choice of
 of $\Lambda_{\rm max}$ are presented in Sect.\ref{rmdep}.

The resulting two-dimensional power spectra of the RM images are shown in the top-left
 panel of Fig.~1. They have the same slope as their parent's three-dimensional magnetic 
field power spectra and they cover an equivalent range of spatial scales. However,
it is evident from Fig.~1 that the same cluster magnetic field energy
density will generate different magnetic field configurations with 
correspondingly different RM structures, for different values of the power
spectrum spectral index.

\subsection{RM radial profiles}
\label{rmprofiles}
RM images of a whole cluster, such as those shown in Fig.~1, cannot be observed 
in real cases. However, it is relatively easy to measure the RM
dispersion and mean (\srm~and \rmm) in limited regions by observing
radio sources located at different projected distances from the center
of a cluster.  The \srm~and \rmm~ values found in radio sources,
 are usually calculated on their project surface which is typically about $50\times 50$
kpc$^{2}$ in size. Since we need to compare the simulation results with these 
 observables, it is necessary to calculate the simulated values of \srm~and \rmm~
 in regions with the same projected area.

Fig.~2 shows the profiles of \srm~(left column panels), 
\absrmm~(central column panels) and \absrmm/\srm~(right column panels)
 as a function of the projected distance from the cluster center, calculated in boxes 
of $50\times 50$ kpc$^{2}$ in size. Due to the random nature of the simulated magnetic field, the values of 
  \srm~and \absrmm~at a given radial distance vary from box to box. We   
calculated the mean and dispersion of \srm~and \absrmm~values from all the boxes located at the same projected
distance from the cluster center. The lines and the shaded regions plotted in Fig.~2 represent
 the mean and dispersion of \srm~and \absrmm, respectively. 

The size of the sampling region used to trace the RM profiles, falls
inside the power spectrum range scales. Therefore, \srm~is explained
by the fact that the magnetic field fluctuates down to scales
smaller than the size of the sampling regions. The most striking
feature appearing in Fig.~2 is the significant amount of
\rmm~predicted by the simulations for any of the three spectral
indexes adopted. This is due to fluctuations of the cluster magnetic
field on scales comparable or greater than the size of the sampling
regions. This is not an unexpected result, but the numerical simulations 
provide a quantitative evaluation of this effect.

As expected, both \srm~and \absrmm~decrease from the cluster center
outward, and the profiles corresponding to a constant magnetic field
strength are the flattest.  Moreover, the ratio of  \absrmm~ to \srm~as
 a function of distance from the cluster center is constant, i.e. both
quantities have the same average radial gradient.

While both \srm~and \absrmm~increase linearly with the average cluster magnetic 
field strength, \bm, and the central gas density (see Eq.~\ref{rm}),
the ratio \absrmm/\srm~ depends only on the magnetic field power spectrum slope, $n$, 
for a given range of fluctuation scales. This means that the comparison of RM data of 
radio galaxies embedded in a cluster of galaxies with simulated profiles, can be used to infer
the strength and the power spectrum slope of the cluster magnetic field.

\subsection{The dependence of RM on the magnetic field largest scale structure}
\label{rmdep}

To trace RM profiles shown in Fig.\ref{fig2} we adopted the largest magnetic field fluctuation scale permitted
 by the computational grid size. In this section we consider different values for $\Lambda_{\rm max}$ and we
 calculate the corresponding variations of the simulated RM.

We performed an additional series of simulations where we decrease the value 
of $\Lambda_{\rm max}$ from 768 kpc down to 6 kpc. We keep the minimum
magnetic field scale $\Lambda_{\rm min}$ fixed at 6 kpc and allowed $\Lambda_{\rm max}$ to vary, 
since, given the limited size of the radio source
RM images, the former can be better estimated from the data with respect to the latter.  

The variation of \srm, \absrmm~and their ratio as a function of $\Lambda_{\rm max}$ is shown 
 in left, middle and right panel of Fig.~3, respectively. 

Note that \srm~and \absrmm~in middle and left panels of Fig.~3 have been normalized to their values
 at $\Lambda_{\rm max}$=768 kpc. Doing this we obtained a scaling factor for the profiles shown in Fig.~2
 which depends only on the power spectrum slope $n$ and on $\Lambda_{\rm max}$ for a fixed $\Lambda_{\rm min}$ . 
We found that for $n$=3, and especially $n$=4, \srm~increases until $\Lambda_{\rm max}$ is larger than the size of
the sampling region in which the statistic is calculated (i.e. $50
\times 50$ kpc$^{2}$) while it stays almost constant for $n$=2.  When
$\Lambda_{\rm max}$ becomes smaller than the sampling region,
\srm~starts to decrease.  Instead, \absrmm~ monotonically decrease by
decreasing $\Lambda_{\rm max}$ and it goes to zero as $\Lambda_{\rm
max}\to \Lambda_{\rm min}$ for all the three spectral indexes.

The right most panel of Fig.~3 shows the behavior of \absrmm/\srm~as a function of $\Lambda_{\rm max}$ 
 for the three different slopes of the magnetic field power spectrum. In general we find the ratio \absrmm/\srm~decreases
 with $\Lambda_{\rm max}$.

If $\Lambda_{\rm max}$ is significantly larger than about 50-100 kpc, a given  value of \absrmm/\srm~can be 
explained by different combinations of parameters: the higher $n$, the lower $\Lambda_{\rm max}$. This could be considered as a 
serious degeneracy at a first glance. However, it is important to bear in mind that the resulting RM images, albeit characterized
 by the same value of \absrmm/\srm, still have very different spatial power spectra and thus different correlation lengths. Therefore, if good RM images are available for the
 cluster radio sources, it is possible to break the degeneracy between $n$ and $\Lambda_{\rm max}$, as will 
be shown for the case of A119 in Sect.~\ref{A119}.

As $\Lambda_{\rm max}$ becomes lower than about 50-100 kpc, the ratio \absrmm/\srm~falls below 0.2-0.3 and its dependence 
on the power spectrum slope vanishes. In this case all the three spectral indexes predict almost the same \absrmm/\srm~which can 
be then used to constrain $\Lambda_{\rm max}$.

In any case, observed values of \absrmm/\srm~greater than unity can only be explained by steep 
magnetic field power spectra. For example, values of \absrmm/\srm$>1$ rule out $n=2$ for any value of $\Lambda_{\rm max}$.

\subsection{ Comparison with single-scale analytical magnetic field model}
\label{monoscale}

The bulk of cluster magnetic field strength estimates appearing in the
literature have been calculated so far by using an analytical
formulation based on the approximation that the magnetic field is
tangled on a single scale $\Lambda_{\rm c}$.  In this formulation, by considering a 
density distribution which follows a $\beta$-profile,
the following relation (hereafter analytical formula) for the RM
dispersion with the  projected distance from the cluster center, $r_{\perp}$, is 
obtained by integrating Eq. 4:

\begin{equation} 
\sigma_{\rm RM}(r_{\perp})= {{K B n_{\rm 0}  r_c^{1/2} \Lambda_{\rm c}^{1/2} }
 \over {(1+r_{\perp}^2/r_c^2)^{(6\beta -1)/4}}} \sqrt {{\Gamma(3\beta-0.5)}\over{\Gamma(3\beta)}}
\label{felten}
\end{equation}

where $\Gamma$ is the Gamma function. 
The constant $K$ depends on the integration path over 
the gas density distribution:
$K$ = 624, if the source lies completely
beyond the cluster, and $K$ = 441 if the source is halfway the cluster.
If the magnetic field scales with the distance from the cluster center 
 as in Eq.~\ref{br}, $\beta$ and $B$ in Eq.~\ref{felten} must be replaced by 
 $\beta+\mu$ and $B_{\rm 0}$, respectively (see Dolag et al. 2001).

It is of interest to compare the
results of our simulations with the predictions of the analytical
approach. However, since in this work we consider a
magnetic field which fluctuates over a range of scales extending at
least one order of magnitude, it is necessary to discuss which is 
the value for $\Lambda_{\rm c}$ we should adopt in Eq.\ref{felten}.

Here we consider three possibilities: i)  $\Lambda_{\rm c}$ is equal to the 
 minimum field scale-length  $\Lambda_{\rm min}$, ii) $\Lambda_{\rm c}$ is 
equal to the  field correlation length\footnote{The magnetic field correlation length is defined as 
\begin{equation}
\Lambda_{\rm B_{z}}=\frac{\int_0^{\infty} w_{\rm z}(r)dr}{w_{\rm z}(0)}
\end{equation}
where $w_{\rm z}(r)$ is the spherically averaged autocorrelation function of the magnetic 
field component along the line-of-sight. The RM correlation length is defined as 
 \begin{equation}
\Lambda_{\rm RM}=\frac{\int_0^{\infty} C_{\rm RM}(r)dr}{C_{\rm RM}(0)}
\end{equation}
where $C_{\rm RM}(r)$ is the radially averaged RM autocorrelation function.
} scale $\Lambda_{\rm B_{\rm z}}$ and 
 iii) $\Lambda_{\rm c}$ is equal to the RM correlation length scale $\Lambda_{\rm RM}$.

\begin{figure}
\begin{center}
\includegraphics[width=10cm]{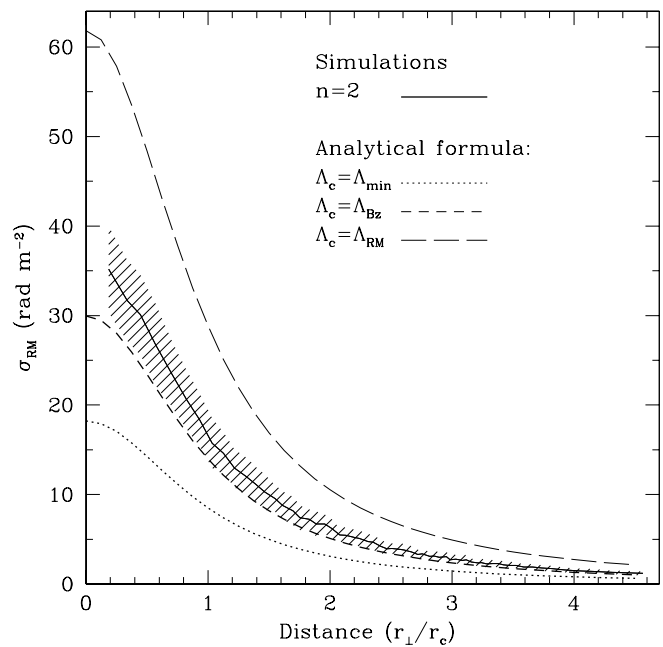}
\end{center}

\caption[] { Comparison of the \srm~profiles obtained from our simulations (solid line) and 
the analytical approximation (dashed and dotted lines). We assumed a source situated halfway through the cluster and the 
same magnetic field strength and gas density radial profiles in both the simulations and the 
analytical formulation. In the simulations we assumed a magnetic field power spectrum with a 
slope $n=2$ and a range of scales from $\Lambda_{\rm min}= 6$ kpc to $\Lambda_{\rm max} = 768$ kpc.}
\label{figfelten}
\end{figure}

In Fig.~\ref{figfelten} we compare the \srm~trend expected from the FARADAY simulations (solid line)
 with the predictions of the analytical formulation for the three scale lengths (dashed and dotted lines).
The simulated trend corresponds to a magnetic field power spectrum with n=2, $\Lambda_{\rm min}= 6$ kpc
 and $\Lambda_{\rm max} = 768$ kpc. This model has a magnetic field correlation length of 
$\Lambda_{\rm B_{\rm z}}\simeq 16$ kpc and a RM correlation length of $\Lambda_{\rm RM}\simeq 69$ kpc. 

High-resolution RM maps of extended radio galaxies
in clusters permit to detect patchy structures on small scales
with typical size in the range 2-15 kpc
(e.g. Feretti et al. 1995, Feretti et al. 1999a, Govoni et al. 2001b,
Taylor et al. 2001, Eilek \& Owen 2002). The cluster magnetic fields strength reported in these 
works has been usually calculated by considering $\Lambda_{\rm c}$ in Eq.~\ref{felten} to be equal to 
the smallest patchy structures detectable in the RM maps. This is the motivation for considering
 $\Lambda_{\rm c}\equiv \Lambda_{\rm min}$. In this case, by assuming the same magnetic field strength,
the single-scale magnetic field approximation leads to a lower \srm~with respect
 to the prediction of our simulations. As a result the magnetic field strength inferred by the 
analytical formula is systematically {\it overestimated} by a factor of about 2
with respect the magnetic field obtained by assuming a multi-scale
magnetic field with a power spectral index of $n=2$.

On the other hand, since for a broad power spectrum $\Lambda_{\rm RM} > \Lambda_{\rm B_{z}}$ 
(see Vogt \& En{\ss}lin 2003), the assumption of the RM correlation length for
 $\Lambda_{\rm c}$ in Eq.~\ref{felten} leads to {\it underestimate} the magnetic field strength
 by the same factor. 

The best agreement between the simulated and analytical \srm~trends is found for $\Lambda_{\rm c}=\Lambda_{\rm B_{z}}$. In 
fact, this should be the proper scale length to use in  Eq.~\ref{felten}. 

To summarize, the analytical formulation leads to a reliable estimate for the average magnetic field strength 
provided that the magnetic field correlation length is known. Hoverer, since $\Lambda_{\rm B_{\rm z}}$ is $n$ dependent, 
we need to estimate the power spectrum of the magnetic field fluctuation in order to use Eq.~\ref{felten}.

\section{Simulated cluster radio sources depolarization}

In this section we investigate the effects of cluster magnetic fields 
 on the depolarization of cluster radio sources.

\subsection{Beam depolarization of cluster radio galaxies}
This kind of depolarization is expected to be particularly strong 
if differential Faraday rotation is occurring within the beam, i.e.
 if the minimum magnetic field scale length is smaller than the beam 
 size.

\begin{figure*}
\begin{center}
\includegraphics[width=15cm]{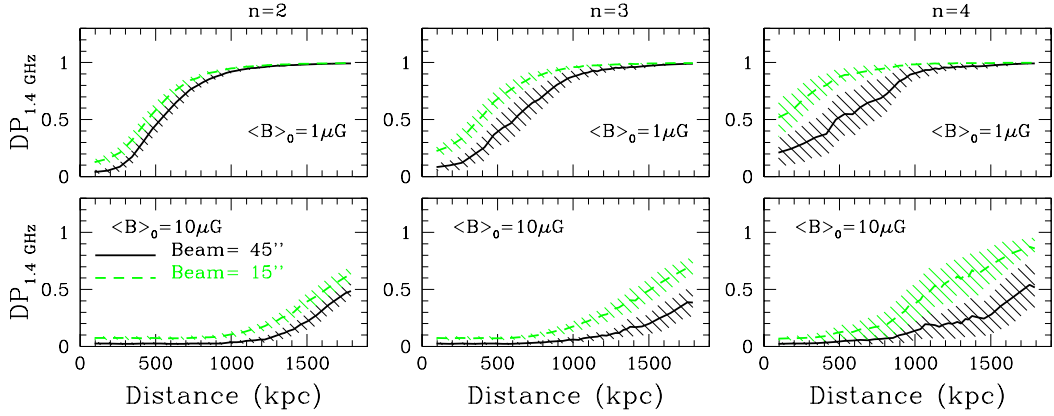}
\end{center}
\caption[]{
Simulated  1.4 GHz beam-depolarization as a function of the distance from the cluster center in the case
 of a varying magnetic field ($\mu=0.3$) with a central strength of \bm$_{\rm 0}=1$ (top panels) and 10 $\mu$G (bottom panels).
 The simulated cluster is placed at a distance of about
 300 Mpc (z=0.05). Solid and dashed lines refer to a beam size of 45\arcsec~and 15\arcsec,
 respectively.
 At the distance of the cluster these beam widths correspond to
projected linear sizes of about 60 and 20 kpc.
}
\label{figdepo}
\end{figure*}

Due to the high RM values in the inner regions of the cluster, the
radio sources located in projection near the cluster center are
expected to have a higher depolarization than those located at the
periphery.  In what follows we refer to the depolarization, DP, as the ratio between the polarization obtained after the 
rotation of the plane of polarization and the intrinsic polarization at a given frequency.  
To study the dependence of this effect on the
cluster magnetic field strength and structure, we simulated the
depolarization of virtual radio sources located  halfway through the cluster at 
increasing distances from the cluster center.
The virtual radio sources are
supposed to lie at the distance of the cluster center and their 
three-dimensional structure and inclination are
neglected.  Their intrinsic polarization intensity, P$_{\rm Int}$, was
considered spatially uniform and completely ordered, i.e. the intrinsic
polarization plane direction, $\Psi_{\rm Int}$, did not vary 
across sources.

We first calculated the expected polarization plane direction, 
$\Psi_{\rm Obs}(\nu)$, 
 at a given frequency along each line-of-sight of the sources
according to  Eq.~\ref{psi}.

We then converted $P_{\rm Int}$ and  $\Psi_{\rm Obs}$ into 
the Stokes parameters $Q(\nu)$ and $U(\nu)$, by inverting the relations:
\begin{eqnarray}
\label{uq}
P&=\sqrt{U^{2}+Q^{2}} \\ \nonumber
\Psi&=\frac{1}{2}\arctan(U/Q)  
\end{eqnarray}

Finally, the simulated $Q(\nu)$ and $U(\nu)$ images have been
 convolved with the desired beam and transformed back to $P_{\rm
 Obs}(\nu)$ and $\Psi_{\rm Obs}(\nu)$ using Eqs.~\ref{uq}.

The parameters of the simulated cluster are those listed in Tab.~\ref{simul}.
In Fig.~\ref{figdepo} we show the depolarization of the sources at 1.4
GHz as a function of the distance from the cluster center for two 
different decreasing ($\mu$=0.3) magnetic field strength profiles with 
 \bm$_{\rm 0}=1$ (top panels) and 10 $\mu$G (bottom panels).

 The simulated cluster
is placed at a distance of about 300 Mpc ($z=0.05$). Solid and
dashed lines refer to a beam size of 45\arcsec~and 15\arcsec,
respectively. At the distance of the cluster these beam widths correspond to
projected linear sizes, $\Lambda_{\rm beam}$, of about 60 and 20 kpc,
i.e.  $\Lambda_{\rm beam}/\Lambda_{\rm min}\simeq 10$ and $\Lambda_{\rm
beam}/\Lambda_{\rm min}\simeq 3$, respectively.

We found that with a beam size of 45\arcsec~the dependence of the
simulated polarization profiles on the adopted power spectrum spectral
index is only marginal, while it becomes significant by considering a beam of 15\arcsec.

From Fig.~\ref{figdepo} it is evident that the depolarization effect is 
 strongly dependent on the cluster magnetic field strength.
In the case of a weak magnetic field strength (e.g. 1
$\mu$G) the Faraday rotation only affects the polarization of the
sources located in projection near the cluster center. As the average
cluster magnetic field strength increases, the beam depolarization
becomes significant at progressively larger distances. For a
 cluster magnetic field with a strength of \bm$_0=10$ $\mu$G
in the center, the sources are severely depolarized for distances up to
 $2\div 3~ r_{\rm c}$ at both 45\arcsec~and 15\arcsec~resolutions.

The comparison of these simulations with the polarization properties of
  large samples of radio sources could provide an additional  
 statistical constraint on the intracluster magnetic field strength.

\subsection{The morphology and (de)polarization of the radio halo emission}
\label{halos}

Cluster radio halos may provide important information about the spatial power
spectrum of the magnetic field fluctuations on large scales ($> 100$
kpc). In particular, different values of the power spectrum spectral
index will generate very different total intensity and polarization
brightness distributions for the radio halo emission.

For a given spectral index $n$, we calculated the expected total
intensity and polarization synchrotron emission by  introducing in
the three-dimensional magnetic field an isotropic population of
relativistic electrons whose distribution follows:
\begin{equation} 
N(\epsilon,\theta)=N_{\rm 0}\epsilon^{-\delta}(\sin\theta)/2 
\label{N}
\end{equation}

where $\epsilon$ and $\theta$ are the electron's energy and the pitch angle between
 the electron's velocity and the direction of the magnetic field, respectively.

At each point of the computation grid, we calculated the 
synchrotron emissivity, $J_{\rm syn}$, by convolving the particle energy distribution
 given in Eq.~\ref{N} with the emission spectrum of the single electron 
\begin{equation}
-\frac{d\epsilon}{dtd\nu}=C_{\rm f} B\sin\theta F\left(\frac{\nu}{C_{\rm \nu}B\sin\theta\epsilon^{2}}\right) 
\end{equation}

where $F(x)$ is the synchrotron radiation kernel 
\begin{equation}
F(x)=x  \int\limits_{x}^{\infty} K_{5/3} (z) dz
\label{fx}
\end{equation}
where $C_{f}$ and $C_{\nu}$ are constants and whose numerical values (cgs units) are $6.26 \times 10^{18}$ and
 $2.34\times 10^{-22}$, respectively, while $K_{5/3}$ is the 5/3-order modified Bessel function.

Let $B_{\perp}$ be the magnetic field perpendicular to the line-of-sight.
Because of the high beaming of the synchrotron radiation, only those electrons with
 $\theta=\sin^{-1}(B_{\perp}/B)$ are visible by the observer.
 Therefore we have
\begin{equation}
 J_{\rm syn}(\nu)=C_{\rm f} \frac{N_{\rm 0}}{2}
 \frac{{B_{\perp}}^{2}}{B} \int\limits_{\epsilon_{\rm min}}^{\epsilon_{\rm max}} F\left (\frac{\nu}{C_{\rm \nu}B_{\perp}\epsilon^{2}}\right) \epsilon^{-\delta} d\epsilon~.
\label{js}
\end{equation}
where $\epsilon_{\rm min}$ and $\epsilon_{\rm max}$ are respectively the low and high energy cut-offs of 
the electron distribution.
We calculated the intrinsic linear polarization emissivity, $P_{\rm int}$, according to
\begin{equation}
P_{\rm int}(\nu)= J_{\rm syn}(\nu) \cdot \frac{\int\limits_{\epsilon_{\rm min}}^{\epsilon_{\rm max}} F\left (\frac{\nu}{C_{\rm \nu}B_{\perp}\epsilon^{2}}\right) \epsilon^{-\delta} d\epsilon} {\int\limits_{\epsilon_{\rm min}}^{\epsilon_{\rm max}} G\left (\frac{\nu}{C_{\rm \nu}B_{\perp}\epsilon^{2}}\right) \epsilon^{-\delta} d\epsilon}
\label{pint}
\end{equation}
where $G(x)=x K_{2/3}(x)$  (see e.g. Rybicki \& Lightman 1979).

At our selected frequencies the radio halo is supposed to be optically thin, therefore we  
calculated the total intensity brightness distribution, $I_{\rm syn}(\nu)$, by simply integrating 
the emissivity in Eq.~\ref{js} along the line-of-sight
\begin{equation}
I_{\rm syn}(\nu)=\frac{1}{4\pi}\int\limits_{0}^{L} J_{\rm syn}(\nu) dl
\label{Is}
\end{equation}
where $L$ is the computational grid size. 

The calculation of the polarization brightness distribution must consider two important
 effects: i) since the direction of  magnetic field changes from point to point inside the cluster
 also the orientation of the intrinsic polarization plane is a spatially varying quantity and ii) 
 as the radio waves propagate through the magnetized intra-cluster medium
 their polarization plane is subject to the {\it internal} Faraday rotation.
 The well known result of this effect is a frequency dependent depolarization of the intrinsic synchrotron emission.

We first calculated the rotated polarization plane direction, $\Psi(\nu,l)$, of the
 emission from  a volume element located at a position $l$ along the line-of-sight ($l$=0 correspond to the farthest side of the computational grid):
\begin{equation}
\Psi(\nu,l)=\Psi_{\rm int}(l)+(c/\nu)^{2} \times {\rm RM}(l)
\label{psiint}
\end{equation}

where $\Psi_{\rm int}(l)$ is perpendicular to $B_{\perp}(l)$ 
and  RM$(l)$ is the internal Faraday rotation occurring in between the volume element and the observer:
\begin{equation}
 {\rm RM}(l)=812\int\limits_l^L n_{\rm e} B_{\|} {\rm d}l~.
\label{rmint}
\end{equation}

It is then convenient again to convert $P_{\rm int}(\nu)$ and $\Psi(\nu,l)$ given in Eqs.~\ref{pint} and \ref{psiint} 
 into the Stokes parameters $Q(\nu,l)$ and $U(\nu,l)$, by inverting the relations given by Eqs.~\ref{uq}.

Finally, we obtained the simulated images of $Q(\nu)$ and $U(\nu)$
 by adding together the contributes of all volumes
along a given line-of-sight
\begin{eqnarray}
Q(\nu)&=\int\limits_{0}^{L} Q(\nu,l) dl \\ \nonumber
U(\nu)&=\int\limits_{0}^{L} U(\nu,l) dl 
\label{uqint} 
\end{eqnarray}

These simulated polarization images take into account of both 
the effects of the field ordering and internal Faraday rotation.

\begin{figure*}
\begin{center}
\includegraphics[width=14cm]{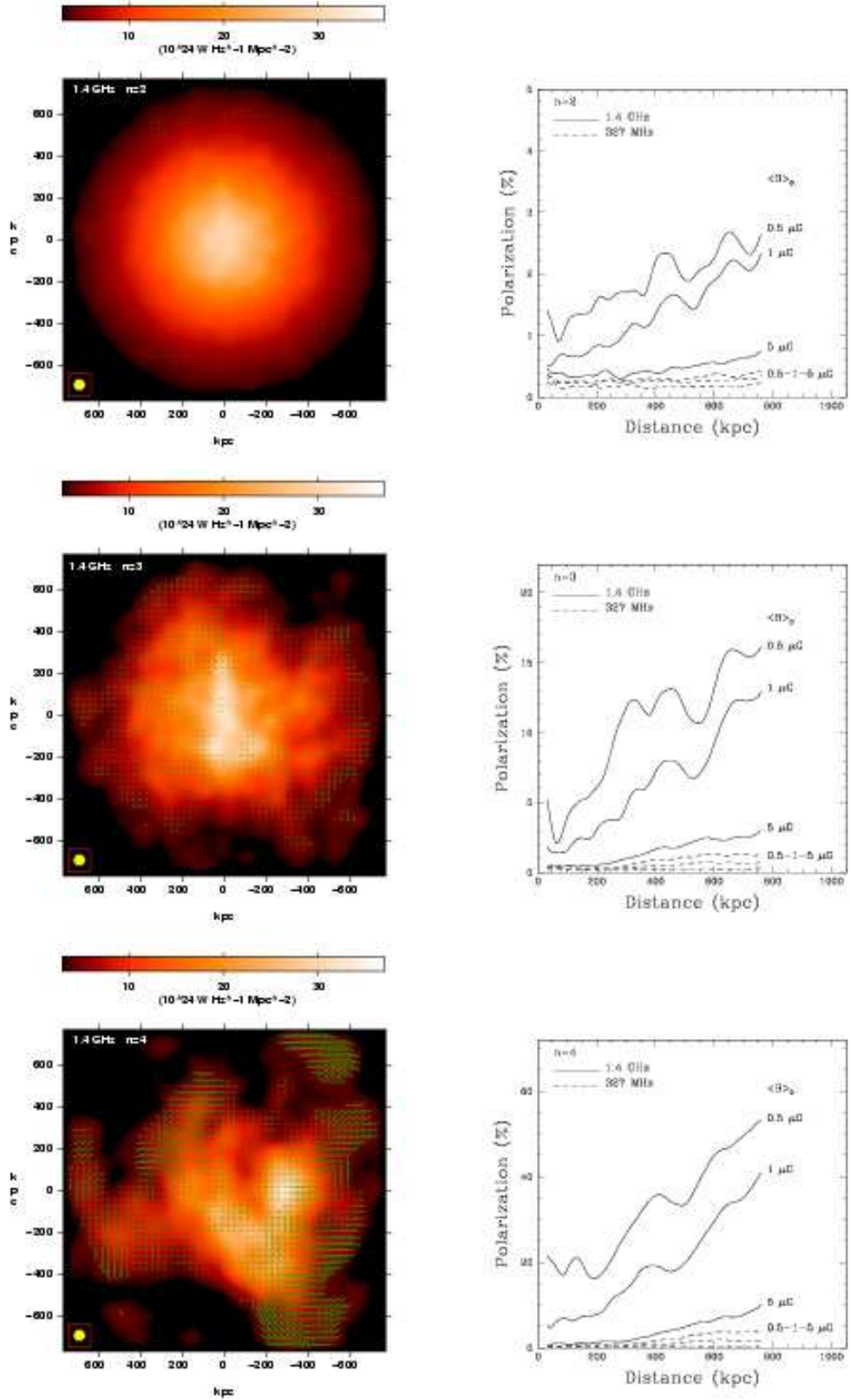}
\end{center}
\caption[]{ simulated synchrotron halo emission for a cluster at a distance of
 $z=0.05$ as it would be observed with a beam of 45\arcsec~($\simeq 60$ kpc).
Left panels: simulated halo brightness and polarization images at 1.4 GHz for different values of the
 magnetic field power spectrum slope $n$ and $\langle B\rangle_{\rm 0}=$1 ${\rm \mu G}$; the vectors lengths are
proportional to the degree of polarization, with 100 percent corresponding
 to 100 kpc on the sky.  Field directions are those of the E-vector. Right panels: radially averaged
 profiles of the polarization percentage at 327 MHz and 1.4 GHz for three values of the magnetic field
strength, namely  $\langle B\rangle_{\rm 0}=$0.5, 1 and 5 ${\rm \mu G}$.}
\label{halobeam}
\end{figure*}
Determining the origin and evolution of the relativistic electron 
 population responsible for the halo emission in cluster of 
galaxy is beyond the purposes of this work. Here we restrict our interest
 to the study the dependence of the radio halo morphology and polarization 
 on the magnetic field strength and structure. We adopted the simplest 
 relativistic electrons distribution as possible which gives 
 plausible values for the halo emission spectral index and total luminosity.

In the simulation we adopt an electron energy spectral index
$\delta=3$.  We choose a Lorentz factor corresponding to the
high-energy cutoff of the electron distribution, $\gamma_{\rm
max}=\epsilon_{\rm max}/(m_{\rm e} c^{2})$, in a way that the
high-frequency cutoff in the emission spectrum (at the cluster
center) occurs at a frequency of 5 GHz:
\begin{equation}
\gamma_{\rm max}=3.4\times10^{4} ~\langle B \rangle_{\rm 0~\mu G}^{-0.5} ~\nu_{\rm 5 GHz}^{~0.5}~.
\label{gammamax}
\end{equation}
The low-energy cutoff, $\gamma_{\rm min}$, and the relativistic electron density, $N_{\rm 0}$,  
 are adjusted to guarantee  a typical average halo brightness at 1.4 GHz of $I_{\rm 1.4}\simeq 5\times 10^{24}$ W Hz$^{-1}$Mpc$^{-2}$ (e.g. Giovannini et al. 1999)
 and a perfect equipartition between the relativistic electron and the magnetic field
 energy densities:
\begin{equation}
 \gamma_{\rm min} \simeq 1.9 \times 10^{3} \langle B \rangle_{\rm 0~\mu G}^{-2}.
\label{gammamin}
\end{equation}

The relativistic electron density $N_{\rm 0}$ follows the magnetic field energy density 
decrease so that the equipartition is satisfied at each distance from the cluster center.

We simulated the expected total  intensity and polarization brightness
 distribution at 1.4 GHz and 
327 MHz, for three values of the magnetic field strength, namely  $\langle B\rangle_{\rm 0}=$0.5, 1 and 5 ${\rm \mu G}$.

Due to the assumed declining ($\mu=0.3$) mean magnetic field strength, the 
spectral index systematically increases from $\alpha_{\rm 1.4 GHz}^{\rm 327 MHz}\simeq 1$ up to  $\alpha_{\rm 1.4 GHz}^{\rm 327 MHz}\simeq 1.3$
moving from the cluster center outward.

 So far, polarization emission from radio halos has not been
detected. The current upper limits to the
  polarization at 1.4 GHz are a few percent ($3-4$\%) for
beams of about 45\arcsec. Due to the low spatial resolution of these
observations, the beam depolarization effect is expected to be very strong.
 Therefore, we simulated the polarization intensity expected at 
 45\arcsec~resolution for a radio halo located at a redshift $z=0.05$.
 At this distance, the corresponding linear resolution is of about 60 kpc. 
In the left-column panels of Fig.~\ref{halobeam} we show the simulated radio halo brightness and 
 polarization percentage distributions at 1.4 GHz  for three different values of the power spectrum 
spectral index $n$. In the right-column panels of  Fig.~\ref{halobeam} we show the expected fractional 
polarization profiles at 1.4 GHz and 327 MHz for the different values of the average magnetic field 
strength.

As one would expect, we found that the degree of polarization percentage:
\begin{itemize}
\item[i)] increases with increasing values of $n$;
\item[ii)] increases with the frequency $\nu$;
\item[iii)] decreases towards the cluster center;
\item[iv)] decreases with increasing magnetic field strength.
\end{itemize}

Our results indicate that a power spectrum slope steeper than $n=3$ and
 a magnetic field strength lower than $\sim 1 {\rm \mu G}$
 result in a radio halo polarization percentage at a frequency of 1.4 GHz
that is far in excess the current observational upper limits at 45\arcsec~
resolution. This means that either the power spectrum spectral index is flatter
 than $n=3$ or the magnetic field strength is significantly higher than
$\sim 1 {\rm \mu G}$. The halo depolarization at 327 MHz is 
 particularly severe and the expected polarization percentage
 at this frequency is always below 1\%.

We also found that the magnetic field power
spectrum slope has a  significant effect in shaping the radio halo.  In
particular, flat power spectrum indexes ($n<3$) give raise to
very smooth radio brightness images (under the assumption that the
radiating electrons are uniformly  distributed). RM 
images of radio galaxies in conjunction with observations of the 
halo brightness and polarization
distributions could provide information  on the behavior of 
 the non-thermal (magnetic field and relativistic
 electrons) components of the intra-cluster medium
 over a large range of spatial scales. For this reason, it is important
to apply our approach to those cluster with both types
of radio sources, such as A2255, in order to determine if a single
magnetic field power spectrum can account for both the observed RM
and radio halo images (Govoni et al. 2002, Govoni et al. 2004 in preparation).
     
\section{Application to the data}

\subsection{The case of Abell 119}
\label{A119}

The cluster of galaxy A119 is an ideal study case to apply our new
approach.  It is characterized by three extended radio galaxies,
0053-015, 0053-016 and 3C29, located at different projected distances
from the cluster center, respectively at 170, 453 and 1515 kpc (see
Tab.~\ref{rmdata} and Fig.~\ref{pow}a).  All these sources are highly polarized and they have been
studied in detail with the Very Large Array (VLA) by Feretti et
al. (1999a).  From RM images with a resolution of 3.75\arcsec 
($\simeq$ 4.4 kpc) Feretti et al. (1999a) estimated a tangling
scale of $\sim 10$ kpc and, by using the simple analytical approach
(see Sect.~\ref{monoscale}), they inferred a strength of about $5
\mu$G for an assumed constant intra-cluster magnetic field. 
However, Dolag et al. (2001) by comparing the X-ray surface
brightness with the \srm~of the three radio galaxies (see Sect.~3.1)
estimated that the magnetic field strength in A119 scales as $B\propto
n_{\rm e}^{0.9}$. Taking into account such a varying magnetic field,
the analytical approach yields a central magnetic field of $\sim 7.5 ~\mu$G.

\begin{figure}[b]
\begin{center}
\includegraphics[width=9cm]{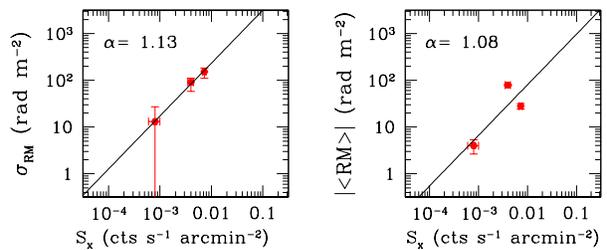}
\end{center}
\caption[]{Correlation of the RM with the X-ray surface brightness 
for the three radio galaxies in A119. We found ${\rm RM}\propto S_{\rm x}^{1.1}$.}
\label{RMSX}
\end{figure}

\begin{figure*}[t]
\begin{center}
\includegraphics[width=16.5cm]{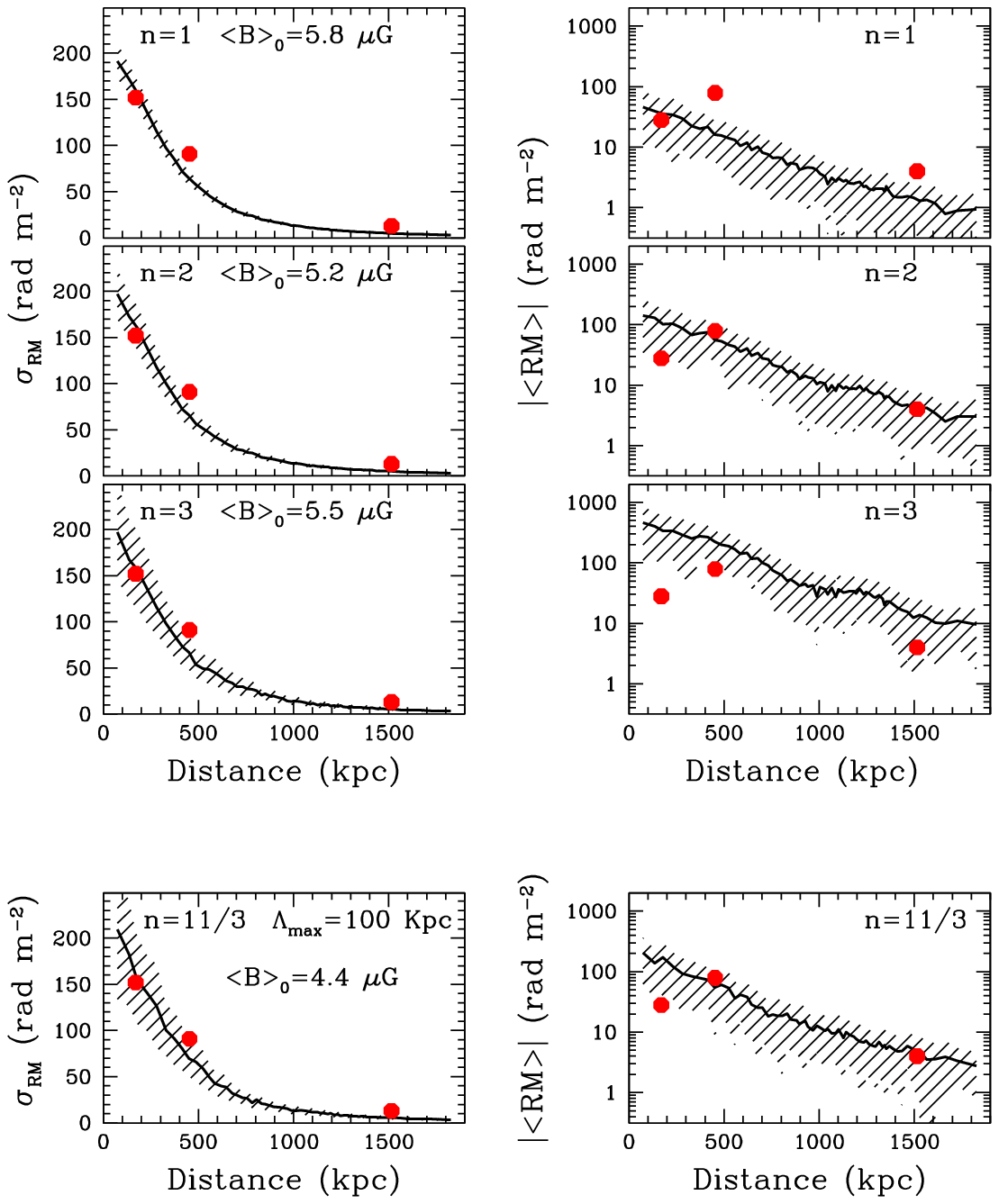}

\end{center}
\caption[] { Best fit of the simulated RM \srm~profiles for the cluster A119 (left column panels).
 The comparison of simulated and observed \absrmm~is shown in right column panels.
 Solid dots refer to the observed values for the radio galaxies 0053-015, 0053-016 and 3C29.
The best fit to the central magnetic field strength gives \bm$_{\rm 0}\simeq 5~\mu$G.
 Since the modeled magnetic field strength decreases from the cluster center outward ($\mu \simeq 0.5$),
 the mean field strength within a spherical volume with radius $r=3\,r_{\rm c}$ is \bm$ \simeq 1.5~\mu$G. We find that the 
spectral index  $n=2$ gives the best description of the data if the magnetic field power spectrum ranges from a minimum scale of 6 kpc up to a maximum
 scale of 768 kpc. The fit corresponding to the models
 with $n \leq 1$ and $n \geq 2$ are definitely poorer since to the \absrmm/\srm~ratios they predict are too low and  too high, 
respectively. The Kolmogorov spectral index ($n=11/3$) gives a reasonable fit to the data only if the maximum 
 scale of the magnetic field fluctuations is about 100 kpc (bottom panels).}
\label{fig8}
\end{figure*}

We used FARADAY  in order to numerically obtain the 
expected RM images  for the
radio galaxies of Abell 119. In the  numerical 
simulations we used the gas
density distribution deduced by X-ray observations for this cluster
($r_{\rm c}=378$ kpc, $n_{0}=1.18\times 10^{-3}~\rm cm^{-3}$ and $\beta=0.56$; Cirimele et al. 1997), and we left the magnetic field power
spectrum slope and strength as free parameters.  All three radio
galaxies are assumed to lie on a plane perpendicular to the
line-of-sight at the distance of cluster center. The
details of the three-dimensional structure of the radio source are
neglected, and the entire observed Faraday rotation is assumed to
occur in the cluster intergalactic medium.

\begin{figure*}
\begin{center}
\includegraphics[width=18cm]{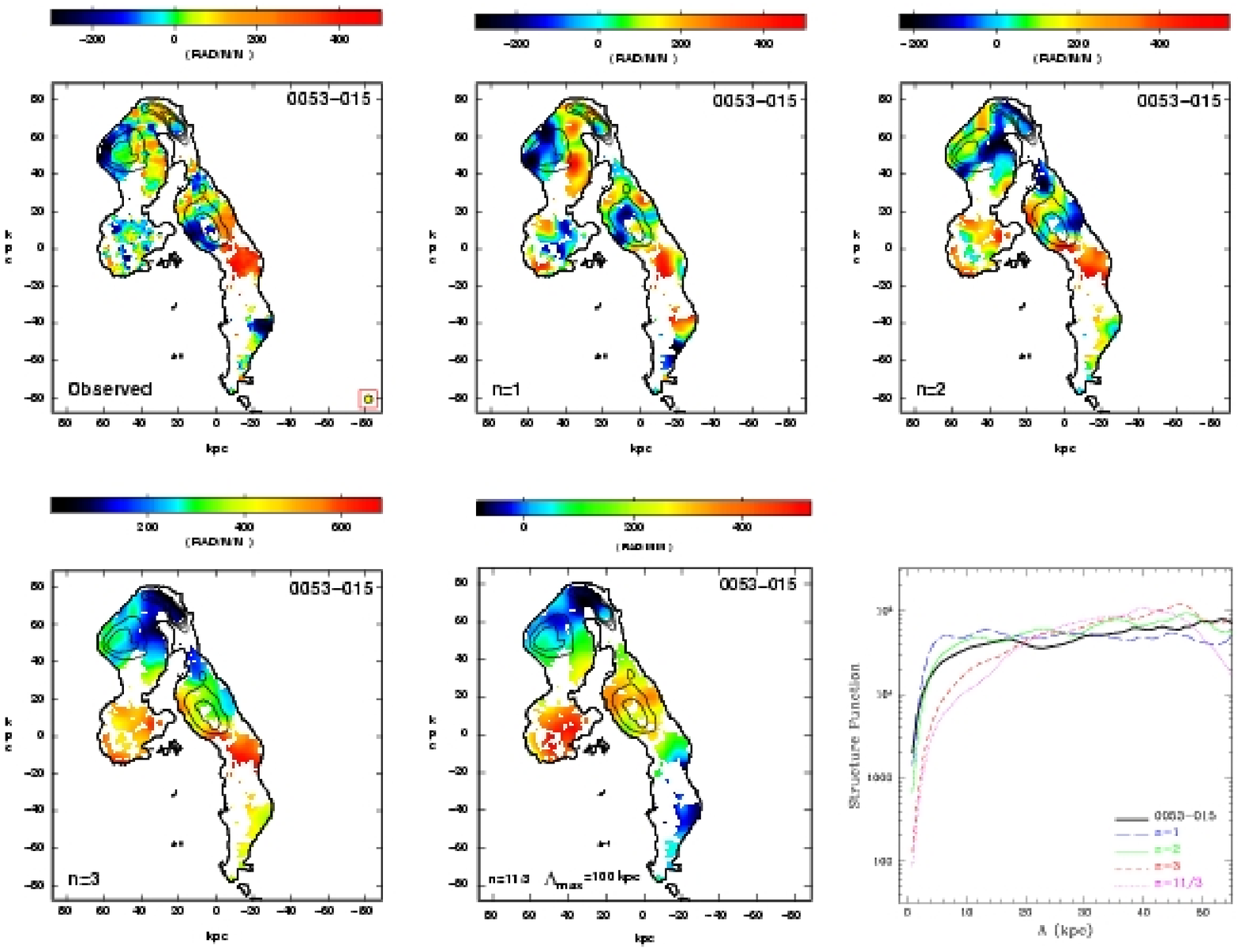}

\end{center}

\caption[] { Comparison of observed and simulated RM images of the source 0053-016 in A119.
 The observed RM image is shown in the top left panel. Models with $n \simeq 2$ better reproduce 
the observed RM features. This is confirmed by the structure functions of the observed and simulated 
RM images shown in the bottom left panel.
}

\label{RM015}
\end{figure*}
 
A119 is located in a region of low Galactic RM (see Tab.~\ref{rmdata}) and
therefore we apply no correction to the observed \rmm~with this
respect. We compared {\it both} \srm~and \absrmm~versus the X-ray
surface brightness $S_{\rm x}$ taken from Dolag et al. (2001) to
estimate the index $\mu$ for the radial scaling of the magnetic
field. Fig.~\ref{RMSX} shows the same dependence ${\rm RM} \propto
S_{\rm x}^{1.1}$ for both quantities, in agreement with the result by
Dolag et al. (2001). From Eq.~\ref{alpha}, with $\beta=0.56$ and
$\alpha=1.1$, it follows that $\mu \simeq 0.5$.
\begin{figure*}
\begin{center}
\includegraphics[width=18.75cm]{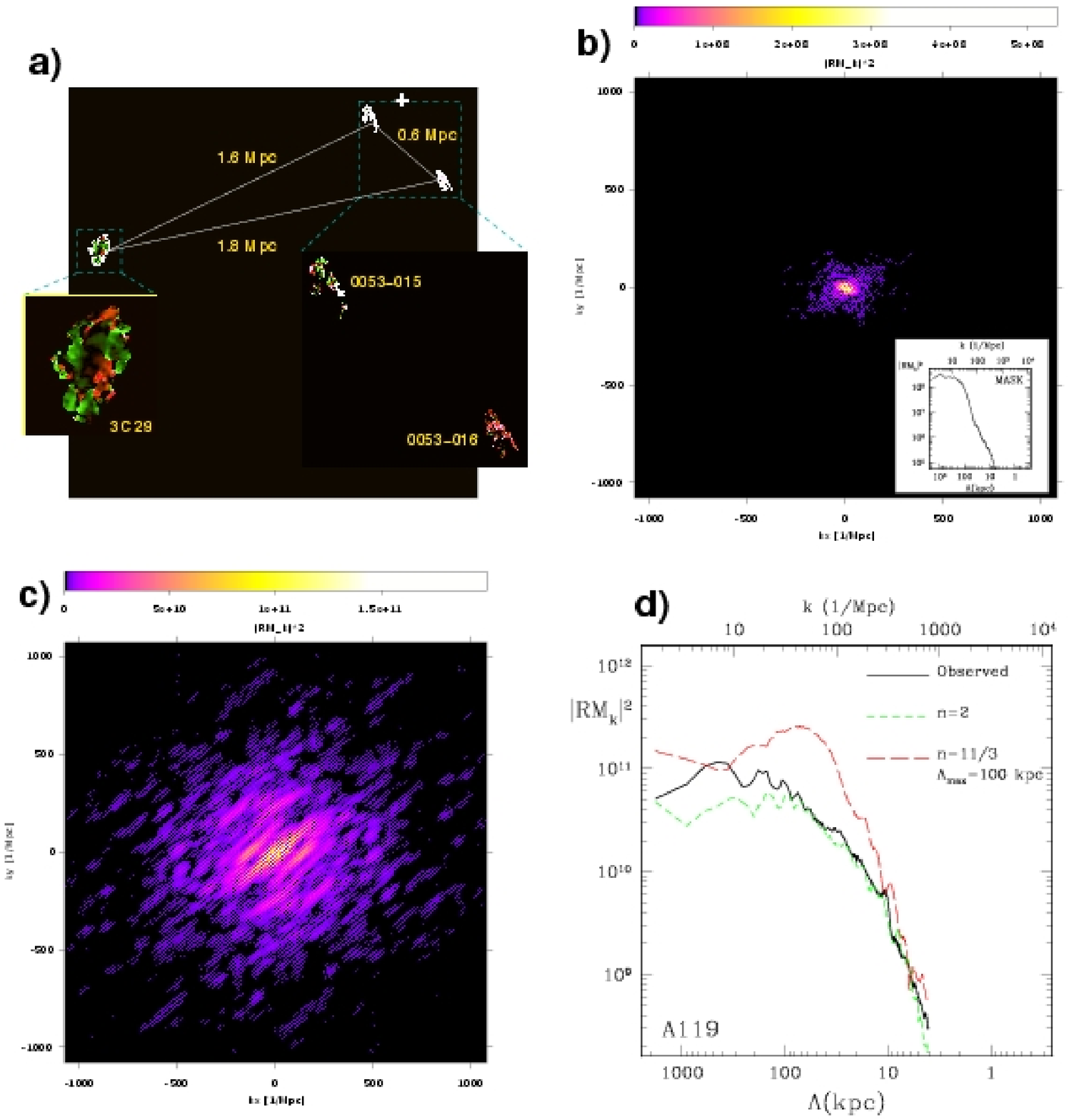}
\end{center}
\caption[] {{\bf a)} Composite RM image of A119 obtained by combining together the individual images of the three radio galaxies. The cross marks the position of the X-ray centroid. {\bf b)} The power spectrum of the mask window function. The short-wavenumber modes are due to the interference between the source-to-source baselines. The inset shows the radially averaged profile of the mask power spectrum. {\bf c)} Power spectrum of the composite RM image. {\bf d)} Comparison of the observed (solid line) and simulated (dashed lines) radially averaged RM power spectra, see text. 
}
\label{pow}
\end{figure*}

As a first step we compared the observed \srm~and \absrmm~with the
expectation of the simulations. Since the linear resolutions  
 of both the computational grid and the observations are very similar,
 we do not apply any Gaussian smoothing to the simulated RM images. 
Fig.~\ref{fig8} show the best fit of
the simulated \srm~radial profiles for the radio galaxies in A119.
The typical uncertainties in the observed RMs are of the order of
10$-$15 rad m$^{-2}$.  However, a realistic error budget should also
consider the effects of a number of systematic uncertainties which are
difficult to estimate, such as the relative distance of the galaxies
from the cluster center along the line-of-sight and the statistical
fluctuations which are intrinsic to the random nature of the magnetic
field. Moreover, the ``observed'' RMs themselves are the result of a
fit procedure as outlined in Sect.~2. For all these reasons, we
decided to use a simplified approach, i.e., a non-weighted least square
fit without attempting to include either \srm~nor \rmm~uncertainties
in the fit procedure.

In the fit procedure we fixed $n$ at a given value and  let
the magnetic field strength as a free parameter.  The best fit of the
magnetic field strength, obtained by fitting the \srm~ profiles
(Fig.~\ref{fig8}, left), gives \bm$_{\rm 0}\simeq 5~\mu$G for each 
$n$. This is the mean magnetic field at the cluster center. Since the
modeled magnetic field strength decreases from the cluster center
outward ($\mu$=0.5), the mean field strength within a spherical
volume of radius $r=3r_{\rm c}$ (with $r_{\rm c}=378$ kpc) 
gives \bm$ \simeq 1.5~\mu$G.  The
comparison between simulated and observed \absrmm~ (Fig.~\ref{fig8},
right) allows us to constrain the value of the index 
 of the power spectrum to $n\simeq2$.  
Fits corresponding to models with $n > 2$ are
definitely poorer as evident from the high \absrmm~values they
predict. Also models characterized by $n < 2$ are a poorer fit because their
low expected \absrmm~values are incompatible with the observations.
As shown in bottom panel of Fig.~\ref{fig8}, the Kolmogorov spectral
index ($n=11/3$) gives a reasonable fit to the data only if the
maximum scale of the magnetic field fluctuations is about 100 kpc.


As expected from the discussion in Sect.\ref{rmdep}, the observed \srm~and the \absrmm~profiles, can be 
explained either by a broad and flat ($n=2$) or by a narrower but steeper ($n=11/3$, $\Lambda_{\rm max}$=100 kpc) 
magnetic field power spectrum.

A way to discriminate the two models is to compare the structure
functions of the observed and simulated RM images of the individual radio galaxies.
As an example, in Fig.~\ref{RM015} we show the observed and simulated RM images relative
to the source nearest to cluster center: 0053-015. The model with
$n=2$ is in remarkable agreement with the observed RM features. This
is confirmed by the structure functions\footnote{ To calculate 
the structure function of the RM images we first construct the
 structure function RM matrix
\begin{equation}
S(dx,dy)=\langle [{\rm RM}(x,y)-{\rm RM}(x+dx,y+dy)]^{2}\rangle
\end{equation}
where $dx$ and $dy$ are both positive and negative offsets and
 the ensemble average is calculated over all the positions, 
 $(x,y)$, in the RM image. The RM structure function
 is then obtained by
 radially averaging $S(dx,dy)$ in annuli of increasing size $\Lambda=\sqrt{dx^2+dy^2}$.}
 plotted in the bottom left panel of Fig.~\ref{RM015}. The observed RM structure function cuts off
below a linear scale of about 10 kpc.  This cutoff and the overall
shape of the observed RM structure functions is reproduced well by
the model with $n=2$. We note that even if the magnetic field power
spectrum with the Kolmogorov index $n=11/3$ and $\Lambda_{\rm
max}=100$ fits the observed RM profiles, it generates a structure
function which is not in agreement with the data. We found similar
results for the other two sources (not shown).

\subsubsection{The composite RM image}

To confirm the above results we calculated the two-dimensional power spectrum of a composite RM image of A119 
 obtained by combining all together the data of the three radio galaxies of the cluster (Fig.\ref{pow}a). 
In Fig.~\ref{pow}b we shown the power spectrum of the ``mask'', i.e. a window function image which is equal to unity 
 if the line-of-sight intercepts one of the sources and zero otherwise. The power spectrum of the composite
 RM image is shown in Fig.~\ref{pow}c. This is the convolution of the true power spectrum with the power spectrum 
of the mask. Since the power spectrum of the mask is quite broad, we cannot measure the power spectrum slope 
directly from the image shown in Fig.~\ref{pow}c. However, we can multiply by the same mask the simulated RM image corresponding 
 to the models with $n=2$ and $n=11/3$ with $\Lambda_{\rm max}=100$, calculate their power spectra, and compare
 them with the observed one. The radially averaged profiles of the observed (solid line) and simulated (dashed lines) 
power spectra of the composite RM image are shown in Fig.\ref{pow}d. Again, we find that the model with $n=2$ gives 
the best description of the data.

To summarize, the best fit of both the RM profiles, structure
functions and power spectra observed for the three extended radio galaxies in A119 is
obtained for a spectral index $n=2$, $\Lambda_{\rm min}$=6 kpc,
$\Lambda_{\rm max}$=770 kpc and \bm$_{\rm 0}\simeq 5~\mu$G. 
The mean field strength within a spherical
volume of radius $r=3r_{\rm c}$ (with $r_{\rm c}=378$ kpc) 
gives \bm$ \simeq 1.5~\mu$G. The observed
\absrmm/\srm~ratios rule out $n=1$ while models with $n>2$, compatible
with the constraint
 $\Lambda_{\rm max}\simeq 100$ kpc, produce RM
images whose structure functions and power spectra are incompatible with the data.
These results are based on the assumption that the observed Faraday
rotation is produced entirely by a magnetized intra-cluster medium.

\subsection{Variations of the power spectrum spectral index among clusters}
We have seen in the case of A119 how by comparing the simulated RMs
with the observed RMs obtained for radio sources located at different
projected  distances from the cluster center, it is possible 
to gain important information about the magnetic field power
spectrum and strength.
\begin{table*}[t]
\caption{RM data taken from the literature.}
\begin{center}
\begin{tabular}{lcccccc}
\noalign{\smallskip}
\hline                  
\noalign{\smallskip}    
Cluster  &    source          &  Distance &Gal.RM &  \rmm        & \srm        & Ref.\\
         &                    &   (kpc)           & (rad/m$^2$)  & (rad/m$^2$) &   \\ \hline 
Coma     &  NGC4869           &   192             &  -6          & -127         &  181       & 1   \\ \hline 
A119     &  0053-015 &   170     &  -1  & +28     &  152       & 2   \\
         &  0053-016 &   453     &      & -79     &   91       &     \\
         &   3C29    &  1515     &      & +4      &   13       &    \\  \hline 
3C129    &  3C129.1  &   50      &  -10 & +21     &  200       &  3   \\ 
         &   3C129   &   500     &      & -125    &   82       &    \\ \hline 
A514     &  PKS0405-203 &   2098    &  -3  & -19     &   48       &  4   \\
         &  PKS0446-206 &   750     &      & 46      &   47       &    \\ \hline 
A2255    &  B1712+640        &   392     &  -6  & -91     &   78       &  5   \\
         &  B1713+641        &   612     &      & +67     &   59       &    \\  \hline 
Zw0056.9+2636   & NGC326 E-lobe    &   70       & -36        & -16   &   58 &  6\\ 
          & NGC326 W-lobe  &   35      &      & -21     &   60        &  \\  \hline     
A400      & 3C75      &   0       &  4    & -7.6    &  100       & 7 \\ \hline
A2634     & 3C465     &   0       &  -60 &-25      &  120        & 7 \\ \hline
3C449     & 3C449     &   0       & -165 &  -166   &   38        & 8\\
         
\hline
\noalign{\smallskip}
\label{rmdata}
\end{tabular}
\begin{flushleft}

Column 1 : Cluster name;
2: Source name;
3: Source projected distance from cluster center;
4: Galactic RM contribute;
5: RM mean;
6: RM dispersion;
7: References;
\smallskip

References: 1) Feretti et al. 1995; 2) Feretti et al. 1999a; 3) Taylor et al. 2001;
4) Govoni et al. 2001a; 5) Govoni et al. 2003 in prep; 6) Murgia et al. 2001;
7) Eilek \& Owen 2002; 8) Feretti et al. 1999b.
\end{flushleft}
\end{center}

\end{table*}
However, one important question concerns the variation of the magnetic
field power spectrum among clusters of galaxies. To explore this
issue, we collected a sample of nine clusters  and groups of galaxies 
(without cooling flows)
for a total of 14 radio sources for which high-quality RM data are
available in the literature. The cluster sample we chose for this
study is listed in Tab.~\ref{rmdata}.  Since we are interested in the mean and
dispersion of the RM, we included in this analysis only those clusters
hosting radio sources for which high-resolution RM images are
available.

The clusters in our sample are characterized by quite different gas
density profiles, linear size and probably magnetic field strengths.
However, since these parameters affect both \absrmm~and \srm~ in the
same way, it is possible to combine the RM data from different
clusters and use the ratio \absrmm/\srm~to study the variation of the
spectral index $n$ among them from a statistical point of view.

The mean of the RM is a fundamental parameter for our analysis and
therefore the contribution of our own Galaxy to this quantity should
be removed carefully.  We determined the galactic contribution to the
\rmm~for each cluster of our sample according to the values reported
by Simard-Normandin et al. (1981). We use the distance-weighted mean
of the RMs of all sources located within $\pm 15\deg$ of galactic
longitude and latitude, respectively.  The galactic RM contributions are
listed in Tab.~\ref{rmdata} (col. 4).

\begin{figure}
\begin{center}
\includegraphics[width=9cm]{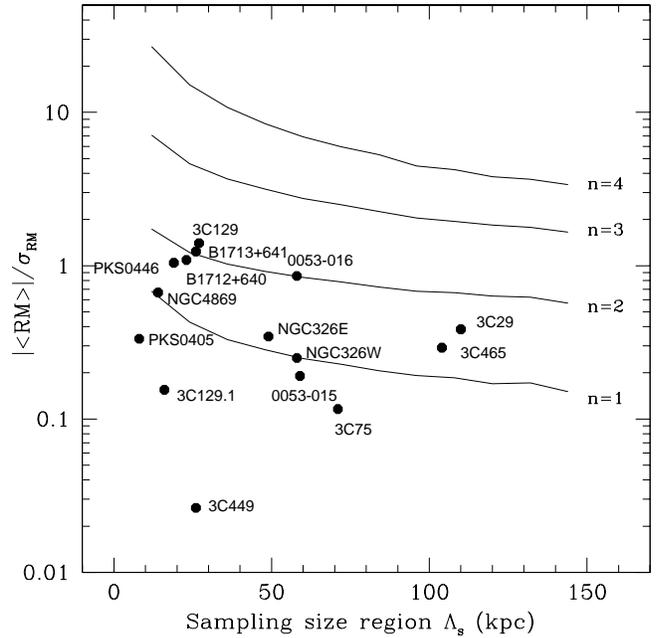}
\end{center}
\caption{The ratio \absrmm/\srm~as a function of the sampling region. 
The horizontal lines represent the simulated trends corresponding to $n$=1, 2, 3 and 4.}
\label{fig11}
\end{figure}

Fig.~\ref{fig11} shows the trend of the observed \absrmm/\srm~ratios
as a function of the sampling region size ($\Lambda_{\rm s}$) for
$n=1$, 2, 3 and 4. The observed values of the \absrmm/\srm~ratio, lie
above 0.1 and below 2 for all clusters except for 3C 449, which has a
very low \absrmm~if corrected for the galactic RM contribution.
 It is important to note that 3C 449 belongs to a small group
and is located at low galactic latitude, therefore
the correction for the galactic RM contribution should be
considered carefully since it could strongly influence
the result. The mean of
\absrmm/\srm~ratio calculated for the 14 radio sources is $\simeq
0.6$.

The values of \absrmm/\srm~expected from the simulations for $n>3$
exceed those observed by a significant amount.  Indeed, as already
found for A119, the rather flat spectral index $n\simeq1-2$ 
represents the best description of the data.  The relatively large 
 dispersion of the observed data is largely consistent with what
 expected from the intrinsic statistical scatter shown in the right 
 panel of Fig.\ref{fig2} (the scatter is larger towards the lower values
 of the \absrmm/\srm~ratio).

 In A2255, for both radio galaxies  the mean of the observed 
RM is greater than its dispersion: \absrmm/\srm$\simeq1.2$.
This is a strong indication
that, at least for this cluster, the magnetic field power spectrum
might be steeper than in the other cases. The inferred  power 
spectrum index could be as high as $n=2$ or more.  A2255 is a very 
interesting cluster since it also contains a wide synchrotron radio halo 
and the magnetic field power spectrum slope has a major role in determining
the large scale morphology of such a diffuse emission (see 
Sect.~\ref{halos}).

\section{Summary}
In this paper we present a numerical approach to investigate the
strength and structure of magnetic fields permeating the intergalactic
medium in clusters of galaxies. 
  The estimates of cluster magnetic field
strengths appearing in the literature have been calculated
 through their effects on the polarization properties 
 of radio galaxies in or behind clusters by using 
 an analytical formulation based on the approximation that
the magnetic field is tangled on a single scale.
However, detailed observations of these radio sources and MHD simulations suggest that
it is necessary to consider cluster magnetic fields which fluctuate
 over a wide range of spatial scales.  To accomplish this we simulated
a random three-dimensional magnetic field with a power law power
spectrum in a cubical box. We calculated the effects of the expected
Faraday rotation on the polarization properties of radio cluster
sources such as radio galaxies and radio halos whose radiation
 crosses the magnetized intracluster medium.
 The main issues of this paper are summarized as follows:

$\bullet$ We determined the statistical behavior of the RM over a
 typical radio galaxy-sized region as a function of distance from the
 cluster center for different values of the magnetic field power
 spectrum slope and range of scales.  The \absrmm/\srm~ratio is an easily
 measurable observable that is expected to reflect the relative amplitude of large to small-scale 
 magnetic field fluctuations. The numerical simulations allow
 a quantitative evaluation of this effect. We found that the \absrmm/\srm~ratio depends rather critically on
 the power spectrum slope, $n$, and therefore it can be used to constrain
 this parameter if spatially-resolved RM images of radio sources
 located at different distances from the cluster center are
 available.

$\bullet$ The magnetic field strength estimated by our simulations in
 the case of $n=2$ (i.e. most of the magnetic field energy resides in
 the small scales) results in a factor of about 2 lower strengths than the
 expectations based on the single-scale magnetic field 
approximation (see Fig. 4). The analytic formulation leads to a reliable 
estimate for the average magnetic field strength only provided that the magnetic field 
correlation length is adopted (i.e. the magnetic field power spectrum is known).

$\bullet$ We simulated the expected dependence of the statistical beam
 depolarization of marginally resolved radio sources located at
 increasing distances from the cluster center. We found that the amount of
 depolarization is strongly dependent on the magnetic field
 strength.

$\bullet$ We point out that radio halos may provide important
 information about the spatial power spectrum of the magnetic field
 fluctuations on large scales.  In particular, different values of the
 power spectrum spectral index give rise to very different total
 intensity and polarization brightness distributions. So far,
 polarization emission from radio halos have not been detected. Our 
results indicate that a power spectrum slope steeper than $n=3$ and 
 a magnetic field strength lower than $\sim 1 {\rm \mu G}$
 predict a radio halo polarization percentage at a frequency of 1.4 GHz
that is far in excess the current observational upper limits at 45\arcsec~
resolution. This means that either the power spectrum spectral index is flatter 
 than $n=3$ or the magnetic field strength is significantly higher than 
$\sim 1 {\rm \mu G}$ or both.

$\bullet$ In the last part of the paper, simulations and data have been
 compared through an interactive approach.  The cluster A119 is an
 ideal case for our new approach.  It is characterized by three
 extended radio galaxies located at different projected distances from
 the cluster center and it is located in a region of very low Galactic
 RM.  The best fit of both the RM profiles and image structure
 functions observed for the three extended radio galaxies in A119 is
 obtained for a spectral index $n=2$ and \bm$_{\rm 0}\simeq
 5~\mu$G. This is the mean magnetic field at the cluster center, the
 mean field strength within a spherical volume with radius $r=3r_{\rm
 c}$ is \bm$ \simeq 1.5~\mu$G. The observed \absrmm/\srm~ratios
 rule out $n=1$ while models with $n>2$, compatible with the
 constraint $\Lambda_{\rm max}\simeq 100$ kpc, produce RM images whose
 power spectra are incompatible with the data.

$\bullet$ An as-yet unexplored issue concerns the variation of the
  magnetic field power spectrum among cluster of galaxies. For this
  purpose, we collected a sample of nine clusters (without cooling
  flows) for a total of 14 radio sources for which high-quality RM data
  are available in literature.  The values of \absrmm/\srm~expected
  from the simulations for $n>3$ exceed the observed ones by a
  significant amount.  Indeed, as already found for A119, the rather
  flat spectral index $n\simeq2$ represents the best description of
  the data.

\begin{acknowledgements}
The authors thank the referee, Dr. De Young, for suggestions that have improved the paper.
We are indebted to Hans-Rainer Kloeckner who provided us the galactic RM tables. We also
 thank Gianfranco Brunetti for stimulating discussions and useful comments.
K. Dolag acknowledge support by a Marie Curie Fellowship of the European
Community program "Human Potential" under contract number MCFI-2001-01227.

\end{acknowledgements}

\end{document}